\documentclass[prd,aps,,preprintnumbers,floats,floatfix,nofootinbib]{revtex4}
\usepackage{graphicx,amsmath}
\usepackage{amssymb}
\usepackage{subfigure}

\begin{document}

\title{The Top Triangle Moose: Combining Higgsless and Topcolor Mechanisms for Mass Generation}

\author{R. Sekhar Chivukula}
\email[]{sekhar@msu.edu}
\author{Neil D. Christensen}
\email[]{neil@pa.msu.edu}
\author{Baradhwaj Coleppa}
\email[]{baradhwaj@pa.msu.edu}
\author{Elizabeth H. Simmons}
\email[]{esimmons@msu.edu}

\affiliation{Department of Physics and Astronomy,\\
Michigan State University,\\ East Lansing, Michigan 48824, USA}

\preprint{MSUHEP-090630}

\date{\today}

\begin{abstract}
We present the details of a deconstructed model that incorporates both Higgsless and top-color mechanisms. The model alleviates the tension between obtaining the correct top quark mass and keeping $\Delta\rho$ small that exists in many Higgsless models. It does so by singling out the top quark mass generation as arising from a Yukawa coupling to an effective top-Higgs which develops a small vacuum expectation value, while electroweak symmetry breaking results largely from a Higgsless mechanism. As a result, the heavy partners of the SM fermions can be light enough
to be seen at the LHC. After presenting the model, we detail the phenomenology, showing that for a broad range of masses, these heavy fermions are discoverable at the LHC.
\end{abstract}

\maketitle

\section{Introduction}

Understanding the mechanism of electroweak symmetry breaking (EWSB) is one of the most exciting problems facing particle physics today. The Standard Model (SM), though phenomenologically successful, relies crucially on the existence of a scalar particle, the Higgs boson \cite{Higgs}, which has not been discovered in collider experiments.
Over the last few years, Higgsless models \cite{Csaki Reference} have emerged as a novel way of understanding the mechanism of EWSB without the presence of a scalar particle in the spectrum. In an extra dimensional context, these can be understood in terms of a $SU(2)\times SU(2)\times U(1)$ gauge theory in the bulk of a finite $AdS$ spacetime \cite{Csaki-Higgsless,Csaki-Higglsess2,Nomura-Higgsless,Sundrum-Higgsless}, with symmetry breaking encoded in the boundary conditions of the gauge fields. These models can be thought of as dual to technicolor models, in the language of the AdS/CFT correspondence \cite{AdS/Cft-1,AdS/Cft-2,AdS/Cft-3,AdS/Cft-4}. One can understand the low energy properties of such theories in a purely four dimensional picture by invoking the idea of deconstruction
\cite{Deconstruction-Georgi,Deconstruction-Hill}. The ``bulk'' of the extra dimension is then replaced by a chain of gauge groups strung together by non linear sigma model fields. The spectrum typically includes extra sets of charged and neutral vector bosons and heavy fermions. The unitarization of longitudinal $W$ boson scattering is accomplished by diagrams involving the exchange of the heavy gauge bosons \cite{Unitarity-1,Unitarity-2,Unitarity-3,Unitarity-4}, instead of a Higgs. A general analysis of Higgsless models \cite{Delocalization-1,Delocalization-2,Delocalization-3,Delocalization-4,Casalbuoni:2005rs,Delocalization-5} suggests that to satisfy the requirements of precision electroweak constraints, the SM fermions have to be `delocalized' into the bulk.
The particular kind of delocalization that helps satisfy the precision electroweak constraints,  ``ideal fermion delocalization'" \cite{IDF}, dictates that the light fermions be delocalized in such a way that they do not couple to the heavy charged gauge bosons. The simplest framework that captures all these ideas, a three site Higgsless model, is presented in \cite{three site ref}, where there is just one gauge group in the bulk and correspondingly, only one set of heavy vector bosons. It was shown that the twin constraints of getting the correct value of the top quark mass and having an admissible $\rho$ parameter necessarily push the heavy fermion masses into the TeV regime \cite{three site ref} in that model.

In this paper, we seek to decouple these constraints by combining the Higgsless mechanism with aspects of topcolor \cite{Hill - Topcolor 1,Hill - Topcolor 2}. The goal is to separate the bulk of electroweak symmetry breaking from third family mass generation. In this way, one can obtain a massive top quark and heavy fermions in the sub TeV region, without altering tree level electroweak predictions. In an attempt to present a minimal model with these features, we modify the three
site model by adding a ``top Higgs'' field, $\Phi,$ that couples preferentially to the top quark.  The resulting model is shown in Moose notation \cite{Moose} in Figure 1; we will refer to it as the ``top triangle moose'' to distinguish it from other three-site ring models in the literature in which all of the links are non-linear sigmal models, such as the ring model explored in \cite{effectiveness} or BESS \cite{BESS-1,BESS-2} and hidden local symmetry \cite{HLS-1,HLS-2,HLS-3,HLS-4,HLS-5} theories.

The idea of a top Higgs is motivated by top condensation models, ranging from the top mode standard model \cite{Nambu:1988mr, Miransky:1988xi, Miransky:1989ds, Marciano:1989mj, Bardeen:1989ds, Marciano:1989xd}  to topcolor assisted technicolor\cite{Hill - TC2,Lane and Eichten - TC2,Hill and Simmons,Popovic:1998vb, Braam:2007pm},  to the top quark seesaw \cite{Dobrescu:1997nm, Top quark seesaw} to bosonic topcolor \cite{Bosonic topcolor,Bosonic topcolor 2}.  The specific framework constructed here is most closely aligned with topcolor assisted technicolor theories \cite{Hill - TC2,Lane and Eichten - TC2,Hill and Simmons,Popovic:1998vb, Braam:2007pm} in which EWSB occurs via technicolor interactions while the top mass has a dynamical component arising from topcolor interactions and a small component generated by an extended technicolor mechanism.    The dynamical bound state arising from topcolor dynamics can be identified as a composite top Higgs field, and the low-energy spectrum includes a top Higgs boson. The extra link in our triangle moose that corresponds to the top Higgs field results in the presence of uneaten Goldstone bosons, the top pions, which couple preferentially to the third generation. The model can thus be thought of as the deconstructed version of a topcolor assisted technicolor model.

We start by presenting the model in section II, and describing the electroweak sector.  The gauge sector is the same as in BESS \cite{BESS-1,BESS-2} or hidden local symmetry \cite{HLS-1,HLS-2,HLS-3,HLS-4,HLS-5} theories, while the fermion sector is generalized from that of the three site model \cite{three site ref} and the symmetry-breaking sector resembles that of topcolor-assisted technicolor  \cite{Hill - TC2,Lane and Eichten - TC2,Hill and Simmons,Popovic:1998vb, Braam:2007pm}.  In section III, we compute the masses and wave functions of the gauge bosons and describe the limits in which we work. We then move on to consider the fermionic sector in section IV. Here, we also explain how the ideal delocalization condition works for the light fermions. In section V, we compute the couplings of the fermions to
the charged and neutral gauge bosons. In section VI, the top quark sector is presented. After calculating the mass of the top quark, we describe how the top quark is delocalized in this model by looking at the tree level value of the $Zb\bar{b}$ coupling. In section VII, we carry out the detailed collider phenomenology of the heavy $U,D,C$ and $S$ quarks. After comparing our phenomenological analysis with others in the literature in section VIII, we present our conclusions in section IX.

\section{The Model}

Before we present the details of our model, we recall the essential features of the closely related three site model that \cite{three site ref} pertain to the heavy fermion mass. The three site model is a maximally deconstructed version of a Higgsless extra dimensional model, with only one extra $SU(2)$ gauge group, as compared to the SM. Thus, there are three extra gauge bosons, which contribute to unitarizing the $W_{L}W_{L}$ scattering in place of a Higgs. The LHC phenomenology of these extra vector bosons is discussed in \cite{Gauge boson phenomenology,Ohl:2008ri}. Also incorporated in the three site model is a heavy Dirac partner for every SM fermion. The presence of these new fermions, in particular, the heavy top and bottom quarks, gives rise to new one-loop contributions to $\Delta\rho$, where $\rho$ is the ratio of the strengths of the low energy isotriplet neutral and charged current interactions. Precision measurements require $\Delta\rho$ to be  $<\mathcal{O}(10^{-3})$ and this constraint, along with the need to obtain the large top quark mass, pushes the heavy quark mass into the multi TeV range, too high to be seen at the LHC. We seek to reduce this
tension by separating the top quark mass generation from the rest of electroweak symmetry breaking in this model, an approach motivated by top-color scenarios.

\begin{figure}[t]
\begin{center}
\includegraphics[width=2in]{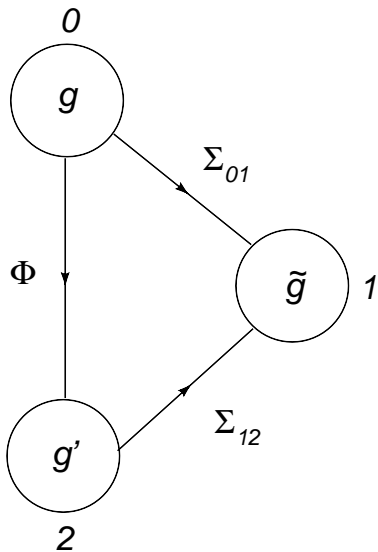}
\caption{The $SU(2) \times SU(2) \times U(1)$ gauge structure of the model in Moose notation \cite{Moose}. The $SU(2)$ coupling $g$ and $U(1)$ coupling $g'$ of sites 0 and 2 are approximately the SM $SU(2)$ and hypercharge gauge couplings, while the $SU(2)$ coupling 
$\tilde{g}$ represents the 'bulk' gauge coupling. The left (right) handed light fermions are mostly localized at site 0 (2) while their heavy counterparts are mostly at site 1. The links connecting sites 0 and 1 and sites 1 and 2 are non linear sigma model fields while the one connecting sites 0 and 2 is a linear sigma field.
}
\label{fig:Triangle}
\end{center}
\end{figure}

The electroweak gauge structure of our model is $SU(2)_0\times SU(2)_1\times U(1)_2$.  This is shown using Moose notation \cite{Moose} in Figure \ref{fig:Triangle}, in which the $SU(2)$ groups are associated with sites 0 and 1, and the $U(1)$ group is associated with site 2.\footnote{Note that
$U(1)_2$ is embedded as a gauged $\tau^3$ of $SU(2)$ -- see Eqn. (\protect\ref{eq:sigma12d}) below.} The SM fermions deriving their $SU(2)$ charges mostly from site 0 (which is most closely associated with the SM $SU(2)$) and the bulk fermions mostly from site 1. The extended electroweak gauge structure of the theory is the same as that of the BESS models \cite{BESS-1,BESS-2}, motivated by models of hidden local symmetry (with $a\neq1$) \cite{HLS-1,HLS-2,HLS-3,HLS-4,HLS-5}.

The non linear sigma field $\Sigma_{01}$ is responsible for breaking the $SU(2)_0\times SU(2)_1$ gauge symmetry down to $SU(2)$, and field $\Sigma_{12}$ is responsible for
breaking $SU(2)_1 \times U(1)_2$ down to $U(1)$. The left handed fermions are $SU(2)$ doublets residing at sites 0 ($\psi_{L0}$) and 1 ($\psi_{L1}$), while the right handed fermions are a doublet under $SU(2)_{1}$($\psi_{R1}$) and two $SU(2)$-singlet fermions at site 2 ($u_{R2}$ and $d_{R2}$). The fermions $\psi_{L0}$, $\psi_{L1}$, and $\psi_{R1}$ have $U(1)$ charges ($Y$) typical of the left-handed $SU(2)$ doublets in the SM, $+1/6$ for quarks and $-1/2$ for leptons. Similarly, the fermion $u_{R2}$ has a $U(1)$ charge typical for the right-handed up-quarks ($+2/3$) and $d_{R2}$ has the $U(1)$ charge typical for the right-handed down-quarks ($-1/3$); the right-handed leptons would, likewise, have $U(1)$ charges corresponding to their SM hypercharge values. Also, the the third component of isospin, $T_3$, takes values $+1/2$ for ``up'' type fermions and $-1/2$ for ``down'' type fermions, just like in the SM. The electric charge assignment follows the relation $Q=T_{3}+Y$. The fermion charge assignments of the quarks
are summarized in Table \ref{tab:chargeswredux}; leptons follow a similar pattern.

\begin{table}[bt]
\begin{center}
\begin{tabular}{|c||c|c|c|}
\hline\hline &&&\\
& \ $\psi_{L0}$,\ & $\psi_{L1}$, $\psi_{R1}$ & $u_{R2}$, $d_{R2}$ \\ [2mm]
\hline\hline &&&\\
$SU(2)_0$& \bf{2}& \bf{1}& \bf{1}\\[2mm]
\hline &&&\\
$SU(2)_1$& \bf{1}&  \bf{2}& \bf{1}\\ [2mm]
\hline &&&\\
$U(1)_2$ & $ \frac16 $ & $\frac16$ & $\frac23$ or $-\frac13$ \\ [3mm]
\hline\hline
\end{tabular}
\caption{Electroweak charge assignments of the quarks.  Leptons follow
a similar pattern. \label{tab:chargeswredux}}
\end{center}
\end{table}

We add a `top-Higgs' link to separate the top quark mass generation from the rest of electroweak symmetry breaking. To this end, we let the top quark couple preferentially to the top Higgs link via the Largangian:
\begin{equation}
\mathcal{L}_{top}=-\lambda_{t}\bar{\psi}_{L0}\,\Phi\, t_{R}+h.c.\label{top quark mass L}
\end{equation}
The top Higgs field is described by the Lagrangian:
\begin{equation}
\mathcal{L}_{\Phi}=\frac{1}{2}D_{\mu}\Phi^{\dagger}D_{\mu}\Phi-V(\Phi),
\label{top Higgs L}
\end{equation}
where the potential $V(\Phi)$ is minimized at $\left<\Phi\right>=f$. When the field $\Phi$ develops a non zero vacuum expectation value, Eqn.(\ref{top quark mass L}) generates a top quark mass term. Since we want most electroweak
symmetry breaking to come from the Higgsless side, we choose the vacuum expectation value associated with the non linear sigmal model fields to be $F=\sqrt{2}\,v$ $\textrm{cos}\,\omega$ (for simplicity, we choose the vev of both the non linear sigma model fields to be the same) and the one associated with the top Higgs sector to be $f=\langle\Phi\rangle=v$ $\textrm{sin\,}\omega$ (where $\omega$ is a small parameter). The top Higgs sector also includes the uneaten Goldstone bosons, the top pions. The interactions of these top pions can be derived from Eqn.(\ref{top Higgs L}) by writing the top Higgs field in the form:
\begin{equation}
\Phi= \left( \begin{array}{c}
(f+H+i\pi^0)/\sqrt{2} \\
i\pi^{-} \end{array} \right),
\end{equation}
where $f$ is defined above and $H$ is the top Higgs. The Extended Technicolor \cite{Dimopoulos:1979es,Eichten:1979ah} induced ``plaquette'' terms that align the technicolor vacuum with the topcolor vacuum and give mass to the top pions can be written as:
\begin{equation}
\mathcal{L}_{\pi}=4 \pi \kappa v^{3} \textrm{Tr}\left({\cal M} \Sigma_{01} \Sigma_{12}^{\dagger}\right),
\end{equation}
where $\kappa$ is a dimensionless parameter and ${\cal M}=(i\sigma^2 \Phi^*\ \Phi)$
is the Higgs field in $2 \times 2$ matrix form. In this paper we restrict our attention to the phenomenology of the fermion and gauge sectors and, therefore, here we assume that the 
top-pions are sufficiently heavy so as not to affect electroweak phenomenology. The phenomenology of the top pion sector will be considered in a future publication.

The mass terms for the light fermions arise from Yukawa couplings of the fermionic fields with the non linear sigma fields
\begin{eqnarray}
\mathcal{L} & = & M_{D}\left[\epsilon_{L}\bar{\psi}_{L0}\Sigma_{01}\psi_{R1}+\bar{\psi}_{R1}\psi_{L1}+\bar{\psi}_{L1}\Sigma_{12}\left(\begin{array}{cc}
\epsilon_{uR} & 0\\
0 & \epsilon_{dR}\end{array}\right)\left(\begin{array}{c}
u_{R2}\\
d_{R2}\end{array}\right)\right].
\label{eqn:Light fermion mass}
\end{eqnarray}
We have denoted the Dirac mass (that sets the scale of the heavy fermion mass) as $M_D$.  Here, $\epsilon_{L}$ is a parameter that describes the degree of delocalization of the left handed fermions and is flavor universal. All the flavor violation for the light fermions is encoded in the last term; the delocalization parameters for the right handed fermions, $\epsilon_{fR}$, can be adjusted to realize the masses and mixings of the up and down type fermions.\footnote{The model has ``next to minimal" flavor violation \protect\cite{Agashe:2005hk}.} For our phenomenological study, we will, for the most part, assume that all the fermions, except the top, are massless and hence will set these $\epsilon_{fR}$ parameters to zero. We will see in Section VI C that even  $\epsilon_{tR}$ is small, since the top quark's mass is dominated by the top Higgs contribution (see Eqn.(1)). Therefore, the top quark mass does not severely constrain $\Delta\rho$, and correspondingly, there will be none of the tension between the heavy quark mass, $M_{D}$, and one loop contributions to $\Delta\rho$, that exists in the three site model. This enables us to have heavy quarks in this model that are light enough to be discovered at the LHC - we will investigate
their phenomenology in Section VII.

\section{Masses and Eigenstates}

In addition to the SM $\gamma$, $Z$ and $W$ bosons, we also have the heavy partners, $W'$ and $Z'$ because of the extra $SU(2)$ group. The canonically normalized kinetic energy terms of the gauge fields
can be written down in the usual way:

\begin{equation}
\mathcal{L}_{KE}=-\frac{1}{4}F^a_{0\mu\nu}F_{0}^{a\mu\nu}-\frac{1}{4}F^b_{1\mu\nu}F_{1}^{b\mu\nu}-\frac{1}{4}B_{\mu\nu}B^{\mu\nu}.
\label{gauge KE terms}
\end{equation}
\\
In this section, we review the masses and wave functions of the gauge bosons, which are the same as those in the BESS model \cite{BESS-1,BESS-2}.

The masses of the gauge bosons come from the usual sigma model Lagrangian:

\begin{equation}
\mathcal{L}_{gauge}=  \frac{F^{2}}{4}\textrm{Tr}[D_{\mu}\Sigma_{01}^{\dagger}D_{\mu}\Sigma_{01}]+\frac{F^{2}}{4}\textrm{Tr}[D_{\mu}\Sigma_{12}^{\dagger}D_{\mu}\Sigma_{12}]+\frac{1}{2}[D_{\mu}\Phi^{\dagger}D_{\mu}\Phi],
\label{Gauge L}
\end{equation}

\noindent where the covariant derivatives are:
\begin{align}
D_{\mu}\Sigma_{01} &=
\partial _{\mu}\Sigma_{01}+igW_{0\mu} \Sigma_{01}-i\tilde{g}\Sigma_{01}W_{1\mu}  \\
D_{\mu}\Sigma_{12} &=
\partial _{\mu}\Sigma_{12}+i\tilde{g}W_{1\mu} \Sigma_{12}-ig'\Sigma_{12}\,\tau^3\, B_{2\mu},
\label{eq:sigma12d}\\
D_\mu \Phi &= \partial_\mu \Phi + igW_{0\mu} \Phi -\frac{ig'}{2} B_{2\mu} \Phi~,
\end{align}
$W_{0\mu} \equiv W_{0\mu}^a \tau^a,\, W_{1\mu} \equiv W_{1\mu}^a \tau^a$ (where $\tau^a=\sigma^a/2$ are $SU(2)$ generators), and $\Sigma_{01}$ and $\Sigma_{12}$ are 2$\times$2 hermitian matrix fields. We will parametrize the gauge couplings in the following form:
\begin{equation}
g_{0}=\frac{e}{\textrm{sin}\,\theta\,\textrm{cos\,}\phi}\,\,\,\,\,\tilde{g}=\frac{e}{\textrm{sin}\,\theta\,\textrm{sin\,}\phi}\,\,\,\,\, g'=\frac{e}{\textrm{cos}\,\theta}.
\label{eqn:gauge couplings}
\end{equation}
\noindent We will find the mass eigenvalues and eigenvectors perturbatively
in the small parameter $\sin\phi$, which we will call $x$. 

\noindent From the above Lagrangian, one can get the mass matrix for the gauge bosons by working in the unitary gauge ($\Sigma_{01}=\Sigma_{12}=1$) and collecting the coefficients of the terms quadratic in the gauge fields. 

The charged gauge boson mass matrix is thus given by:

\begin{equation}
M_{W}^{2}=\frac{e^{2}\, v^{2}}{4\, x^{2}\,\textrm{sin}^{2}\,\theta}\,\left(\begin{array}{cc}
\frac{x^{2}}{1-x^{2}}(1+\textrm{cos}^{2}\,\omega) & -\frac{2 x}{\sqrt{1-x^{2}}}\textrm{cos}^{2}\,\omega \\
-\frac{2 x}{\sqrt{1-x^{2}}}\textrm{cos}^{2}\,\omega & 4\,\textrm{cos}^{2}\omega\end{array}\right).
\label{W mass matrix}
\end{equation}
\noindent This matrix can be diagonalized perturbatively in $x$. We find the light $W$ has the following mass and eigenvector (note that the above formulae are valid to corrections of $\mathcal{O}(x^{3})$, as are all the other eigenvalues and couplings in this paper):

\begin{equation}
M_{W}^{2}=\frac{e^{2}v^{2}}{4\,\textrm{sin}^{2}\,\theta}\left(1+\frac{3x^{2}}{4}\right)
\label{W mass}
\end{equation}

\begin{equation}
W^{\mu}  =  v_{w}^{0}W_{0}^{\mu}+v_{w}^{1}W_{1}^{\mu}=\left(1-\frac{x^{2}}{8}\right)W_{0}^{\mu}+\frac{1}{2}xW_{1}^{\mu}.
\label{W eigen vector}
\end{equation}

\noindent Here, $W_{0}$ and $W_{1}$ are the gauge bosons associated with sites 0 and 1. Since $x$ is small, we note that the light $W$ resides primarily at site 0.  The heavy $W$ eigenvector is orthogonal to the above and has a mass:

\begin{equation}
M_{W'}^{2}=\frac{e^{2}\, v^{2}\textrm{cos}^{2}\,\omega}{4\,\textrm{sin}^{2}\,\theta\, x^{2}}\left(4+x^{2}\right).
\label{W' mass}
\end{equation}

\noindent To leading order, the relation between the light and heavy charged gauge boson masses is

\begin{equation}
\frac{M_{W}^{2}}{M_{W'}^{2}}=\frac{x^{2}}{4\,\textrm{cos}^{2}\,\omega}.
\label{ratio of W masses}
\end{equation}

The neutral gauge bosons' mass matrix is given by:

\begin{eqnarray}
M_{Z}^{2}=\frac{e^{2}\, v^{2}}{4\, x^{2}\,\textrm{sin}^{2}\,\theta}\left(\begin{array}{ccc}
\frac{x^{2}}{1-x^{2}}(1+\textrm{cos}^{2}\,\omega) & -\frac{2 x}{\sqrt{1-x^{2}}}\textrm{cos}^{2}\,\omega & -\frac{x^{2}}{\sqrt{1-x^{2}}}\textrm{sin}^{2}\omega\,\textrm{tan}\,\theta \\
-\frac{2 x}{\sqrt{1-x^{2}}}\textrm{cos}^{2}\,\omega & 4\,\textrm{cos}^{2}\omega & -2\, x\textrm{\, cos}^{2}\,\omega\textrm{\, tan}\,\theta\\
-\frac{x^{2}}{\sqrt{1-x^{2}}}\textrm{sin}^{2}\omega\,\textrm{tan}\,\theta & -2\, x\,\textrm{cos}^{2}\,\omega\,\textrm{tan}\,\theta & x^{2}(1+\textrm{cos}^{2}\,\omega)\textrm{tan}^{2}\,\theta\end{array}\right).
\label{Z mass matrix}
\end{eqnarray}
This mass matrix has a zero eigenvalue (the photon), the eigenvector
of which may be written exactly as:

\begin{equation}
A^{\mu}=\frac{e}{g}W_{0}^{\mu}+\frac{e}{\tilde{g}}W_{1}^{\mu}+\frac{e}{g'}B^{\mu}.
\label{photon eigenvector}
\end{equation}
Requring that this state be properly normalized, we get the relation between
the couplings implied by Eqn. (\ref{eqn:gauge couplings}):

\begin{equation}
\frac{1}{e^{2}}=\frac{1}{g^{2}}+\frac{1}{\tilde{g}^{2}}+\frac{1}{g'^{2}}.
\label{electric charge}
\end{equation}
The light $Z$ boson has the mass

\begin{equation}
M_{Z}^{2}=\frac{e^{2}\, v^{2}}{4\,\textrm{sin}^{2}\,\theta\,\textrm{ cos}^{2}\,\theta}\left(1+x^{2}\left(1-\frac{\textrm{sec}^{2}\,\theta}{4}\right)\right),
\label{Z mass}
\end{equation}
and the corresponding eigenvector 
\begin{equation}
Z^{\mu} =  v_{z}^{0}W_{0}^{\mu}+v_{z}^{1}W_{1}^{\mu}+v_{z}^{2}B^{\mu},
\label{Z eigenvector} 
\end{equation}
where
\begin{equation}
v_{z}^{0}  =  \frac{1}{8}\,(4(-2+x^{2})\textrm{cos\,}\theta-3\, x^{2}\textrm{sec}\,\theta), \,\,\,\,
v_{z}^{1}  =  \frac{1}{2}\, x(-2\textrm{cos}^{2}\,\theta+1)\textrm{sec}\,\theta, \,\, \,\,\,\,
v_{z}^{2}  =  \textrm{sin}\,\theta-\frac{1}{2}x^{2}\,\textrm{sec}\,\theta\,\textrm{tan}\,\theta. \nonumber
\end{equation}
The heavy neutral vector boson, which we call $Z'$, has a mass and eigenvector

\begin{equation}
M_{Z'}^{2}=\frac{e^{2}\, v^{2}\,\textrm{cos}^{2}\,\omega}{4\,\textrm{sin}^{2}\,\theta\, x^{2}}\left(4+x^{2}\textrm{sec}^{2}\,\theta\right)
\label{Z' mass}
\end{equation}

\begin{equation}
Z'^{\mu}  =  v_{z'}^{0}W_{0}^{\mu}+v_{z'}^{1}W_{1}^{\mu}+v_{z'}^{2}B^{\mu},
\label{Z' eigenvector}
\end{equation}
where
\begin{equation}
v_{z'}^{0}  =  \frac{1}{2}\, x, \,\,\,\,
v_{z'}^{1}  =  -1+\frac{1}{8}\, x^{2}\,\textrm{sec}^{2}\,\theta, \,\, \textrm{and} \,\,\,\,
v_{z'}^{2}  =  \frac{1}{2}\, x\textrm{\, tan}\,\theta. \nonumber
\end{equation}
For small $x$, it is seen that the $Z'$ is mainly located at site 1, while
the $Z$ is concentrated at sites 0 and 2, as one would expect.

\section{Fermion wave functions and Ideal delocalization}

In this section, we will review the masses and wave functions of the light fermions and their heavy partners. We will then discuss how to ``ideally delocalize'' the light fermions, which will make the tree level value of the $S$ parameter vanish \cite{IDF}.

\subsection{Masses and wave functions}

Working in the unitary gauge ($\Sigma_{01}=\Sigma_{12}=1),$ the mass matrices of the light quarks and their heavy partners can be derived from Eqn. (\ref{eqn:Light fermion mass}) and take the form:
\begin{eqnarray}
M_{u,d}=M_{D}\left(\begin{array}{cc}
\epsilon_{L} & 0\\
1 & \epsilon_{uR,dR}\end{array}\right).
\label{fermion mass matrix}
\end{eqnarray}
The subscripts $u (d)$ denote up (down) quarks and $M_D$ is the Dirac mass, introduced in Eqn. (\ref{eqn:Light fermion mass}). Diagonalizing the matrix perturbatively in $\epsilon_{L}$, we find
the light eigenvalue:

\begin{equation}
m_{f}=\frac{M_{D}\epsilon_{L}\epsilon_{fR}}{\sqrt{1+\epsilon_{fR}^{2}}}\left[1-\frac{\epsilon_{L}^{2}}{2(1+\epsilon_{fR}^{2})}+...\right].
\label{eqn:light quark mass}
\end{equation}
Note that $m_{f}$ is proportional to the flavor-specific parameter $\epsilon_{fR}$, where $f$ is any light SM fermion (except the top).
The heavy Dirac quark has a mass:

\begin{equation}
m_{F}=M_{D}\sqrt{1+\epsilon_{fR}^{2}}\left[1+\frac{\epsilon_{L}^{2}}{2(1+\epsilon_{fR}^{2})^{2}}+....\right].
\label{eqn:heavy quark mass}
\end{equation}
The left and right handed eigenvectors of the light up quarks are

\begin{eqnarray}
u_{L} & = & u_{L}^{0}\psi_{L0}+u_{L}^{1}\psi_{L1}=\left(-1+\frac{\epsilon_{L}^{2}}{2(1+\epsilon_{uR}^{2})^{2}}\right)\psi_{L0}+\left(\frac{\epsilon_{L}}{1+\epsilon_{uR}^{2}}\right)\psi_{L1}
\label{uL vector}
\end{eqnarray}

\begin{eqnarray}
u_{R} & = & u_{R}^{1}\psi_{R1}+u_{R}^{2}u_{R2}=\left(-\frac{\epsilon_{uR}}{\sqrt{1+\epsilon_{uR}^{2}}}+\frac{\epsilon_{L}^{2}\epsilon_{uR}}{(1+\epsilon_{uR}^{2})^{5/2}}\right)\psi_{R1}+\left(\frac{1}{\sqrt{1+\epsilon_{uR}^{2}}}+\frac{\epsilon_{L}^{2}\epsilon_{uR}^{2}}{(1+\epsilon_{uR}^{2})^{5/2}}\right)u_{R2}.
\label{uR vector}
\end{eqnarray}
The left and right handed eigenvectors of the heavy partners (denoted by $U_{L,R}$) are orthogonal
to Eqn.(\ref{uL vector}) and Eqn.(\ref{uR vector}):

\begin{eqnarray}
U_{L} & = & U_{L}^{0}\psi_{L0}+U_{L}^{1}\psi_{L1}=\left(-\frac{\epsilon_{L}}{1+\epsilon_{uR}^{2}}\right)\psi_{L0}+\left(-1+\frac{\epsilon_{L}^{2}}{2(1+\epsilon_{uR}^{2})^{2}}\right)\psi_{L1}
\label{UL vector}
\end{eqnarray}

\begin{eqnarray}
U_{R} & = & U_{R}^{1}\psi_{R1}+U_{R}^{2}u_{R2}=\left(-\frac{1}{\sqrt{1+\epsilon_{uR}^{2}}}-\frac{\epsilon_{L}^{2}\epsilon_{uR}^{2}}{(1+\epsilon_{uR}^{2})^{5/2}}\right)\psi_{R1}+\left(-\frac{\epsilon_{uR}}{\sqrt{1+\epsilon_{uR}^{2}}}+\frac{\epsilon_{L}^{2}\epsilon_{uR}}{(1+\epsilon_{uR}^{2})^{5/2}}\right)u_{R2}.
\label{UR vector}
\end{eqnarray}
The eigenvectors of other fermions
can be obtained by the replacement $\epsilon_{uR}\rightarrow\epsilon_{fR}$.

\subsection{Ideal fermion delocalization}

The \emph{tree level} contributions to precision measurements in Higgsless models come from the coupling of standard model fermions to the heavy gauge bosons and deviations in SM couplings. It was shown in \cite{IDF} that it is possible to delocalize the light fermions in  such a way that they do not couple to these heavy bosons and thus minimize the deviations in precision electroweak parameters. The coupling of the heavy $W'$ to SM fermions is of the form $\sum_{i}g_i(f^i)^2 v_{W'}^i$. Thus choosing the light fermion profile such that $g_i(f^i)^2$ is proportional to $v_{W}^i$ will make this coupling vanish because the $W'$ and $W$ fields are orthogonal to one another. This procedure (called ideal fermion delocalization \cite{IDF}) makes the coupling of the $W$ to two light fermions  equal the SM value to corrections $\mathcal{O}(x^4)$ and keeps deviations from the standard model of all electroweak quantitites at a phenomenologically acceptable level. Thus, an equivalent way to impose ideal fermion delocalization (IFD) is to demand that the tree level $g_{We\nu}$ coupling equal the SM value.

We will use the latter procedure to implement IFD. The deviation of the $g_{We\nu}$ coupling from the SM value can be parametrized in terms of the $S,$ $T$ and $U$ parameters as \cite{Delocalization-4}:
\begin{equation}
g_{We\nu}=\frac{e}{s}\left[1+\frac{\alpha S}{4s^{2}}-\frac{c^{2}\alpha T}{2s^{2}}-\frac{(c^{2}-s^{2})\alpha U}{8s^{2}}\right].
\label{eqn:gSM}
\end{equation}
where $c=\textrm{cos}\,\theta_{w}=M_{W}/M_{Z}$ and $s=\textrm{sin}\,\theta_{w}=\sqrt{1-c^{2}}$
are related to the ``mass defined'' weak mixing angle. It was shown in \cite{Delocalization-1} that at tree level, in models of this kind, $T,U=\mathcal{O}(x^{4})$, and so we can impose ideal fermion delocalization by requiring $S$ to vanish at tree level, which would make $g_{We\nu}$  in this model equal to the SM value, from Eqn. (\ref{eqn:gSM}). 

In computing the couplings, we will use the mass defined angles; we will indicate this by a suffix $w$ in all the couplings. From Eqns.(\ref{W mass}) and (\ref{Z mass}), we can see that $\textrm{sin}\,\theta_{w}$
is related to $\textrm{sin}\,\theta$ defined implicitly in the couplings in Eqn. (\ref{eqn:gauge couplings}) by:

\begin{equation}
\textrm{sin}\,\theta_{w}=\left(1-\frac{x^{2}}{8}\right)\textrm{sin}\,\theta.
\label{mixing angle definition}
\end{equation}
Using the $W$ and the fermion wave functions, we can calculate the coupling $g_{We\nu}$ as
\begin{equation}
g_{We\nu}=\frac{e}{\textrm{sin}\,\theta_{w}}\left(1+\frac{x^{2}}{4}-\frac{\epsilon_{L}^{2}}{8}\right).
\end{equation}
Thus, we determine the ideal fermion delocalization condition in this model
to be:
\begin{equation}
\epsilon_{L}^{2}=\frac{x^{2}}{2},
\label{eqn:IDF}
\end{equation}
which is the same as in the three-site model, to this order.

\section{Light Fermion couplings to the gauge bosons}

\subsection{Charged Currents}

Now that we have the wave functions of the vector bosons and the fermions, we can compute the couplings between these states. Since all the light fermions are approximately massless, we set $\epsilon_{fR}$ for all the light fermions to zero in this section. We will calculate all couplings to $\mathcal{O}(x^{2})$. We begin with the left handed $Wud$ coupling. 

\begin{equation}
g_{L}^{Wud}  =  g_{0}v_{w}^{0}u_{L}^{0}d_{L}^{0}+\tilde{g}v_{w}^{1}u_{L}^{1}d_{L}^{1}=\frac{e}{\textrm{sin}\,\theta_{w}}.
\label{WudL}
\end{equation}
This result follows from the fact that we have implemented ideal fermion
delocalization in the model. 
All other charged current couplings (both left- and right-handed) can be similarly computed and we present the results in Table II.

\renewcommand{\arraystretch}{1.5}
\begin{table}
\begin{center}
\resizebox{11cm}{!} {
  \begin{tabular}{| c || c | c | }
    \hline
     Coupling & computed as & Strength \\ \hline\hline
     $g_{L}^{Wud}$ & $g_{0}v_{w}^{0}u_{L}^{0}d_{L}^{0}+\tilde{g}v_{w}^{1}u_{L}^{1}d_{L}^{1}$ & $\frac{e}{\textrm{sin}\,\theta_{w}}$ \\ \hline
     $g_{L}^{WUd}(=g_{L}^{WuD})$ & $g_{0}v_{w}^{0}U_{L}^{0}d_{L}^{0}+\tilde{g}v_{w}^{1}U_{L}^{1}d_{L}^{1}$ & $\frac{e\, x}{2\,\sqrt{2}\,\textrm{sin}\,\theta_{w}}$\\ \hline
     $g_{L}^{WUD}$ & $g_{0}v_{w}^{0}U_{L}^{0}D_{L}^{0}+\tilde{g}v_{w}^{1}U_{L}^{1}D_{L}^{1}$ & $\frac{e}{2\textrm{\, sin}\,\theta_{w}}\left(1+\frac{3}{8}x^{2}\right)$ \\ \hline
     $g_{R}^{Wud}$ & $\tilde{g}v_{w}^{1}u_{R}^{1}D_{R}^{1}$ & $0$ \\ \hline
     $g_{R}^{WUd} (=g_{R}^{WuD})$ & $\tilde{g}v_{w}^{1}U_{R}^{1}D_{R}^{1}$ & $0$\\ \hline
     $g_{R}^{WUD}$ & $\tilde{g}v_{w}^{1}U_{R}^{1}D_{R}^{1}$ & $\frac{e}{2\,\textrm{sin}\,\theta_{w}}\left(1-\frac{1}{8}x^{2}\right)$ \\ \hline
     $g_{L}^{W'ud}$ & $g_{0}v_{w'}^{0}u_{L}^{0}d_{L}^{0}+\tilde{g}v_{w'}^{1}u_{L}^{1}d_{L}^{1}$ & $0$\\ \hline
     $g_{L}^{W'Ud}(=g_{L}^{W'uD})$ & $g_{0}v_{w'}^{0}U_{L}^{0}d_{L}^{0}+\tilde{g}v_{w'}^{1}U_{L}^{1}d_{L}^{1}$ & $-\frac{e}{\sqrt{2}\textrm{\, sin}\,\theta_{w}}$  \\ \hline
     $g_{L}^{W'UD}$ & $g_{0}v_{w'}^{0}U_{L}^{0}D_{L}^{0}+\tilde{g}v_{w'}^{1}U_{L}^{1}D_{L}^{1}$ & $\frac{e}{x\,\textrm{sin}\,\theta_{w}}\left(1-\frac{3}{4}x^{2}\right)$  \\ \hline
     $g_{R}^{W'ud}$ & $\tilde{g}v_{w'}^{1}u_{R}^{1}d_{R}^{1}$ & $0$ \\ \hline
     $g_{R}^{W'Ud} (=g_{R}^{W'uD})$ & $\tilde{g}v_{w'}^{1}U_{R}^{1}d_{R}^{1}$ & $0$ \\ \hline
     $g_{R}^{W'UD}$ & $\tilde{g}v_{w'}^{1}U_{R}^{1}D_{R}^{1}$ & $\frac{e}{x\,\textrm{sin}\,\theta}_{w}\left(1-\frac{1}{4}x^{2}\right)$  \\ \hline
  \end{tabular}
  }
  \caption{The couplings of the light and heavy quarks with the charged gauge bosons. Ideal fermion delocalization renders the $g^{Wud}$ coupling the same as the SM value. The coupling of the heavy gauge boson to two heavy quarks is seen to be proportional to $1/x$, which makes $\Gamma_{W'}/M_{W'}>1$, for very small $x$.}
\end{center}
\label{tab:charged couplings}
\end{table}

Two comments are in order. The right handed couplings of the $W$ and $W'$ gauge bosons to two light quarks or to one light and one heavy quark are zero in the limit $\epsilon_{fR}=0$, because in this limit the right handed light quarks are localized at site 2, and the charged gauge bosons live only at sites 0 and 1. The non-zero right handed coupling of $W$ with two heavy fields arises, in this limit, solely from site 1. The left and right-handed $W'$ coupling to two heavy fermions is enhanced by a factor $1/x$ relative to $g_{L,R}^{WUD}$, with $x$ being the small expansion parameter. Thus, if $W'$ is massive enough to decay to two heavy fermions, the width to mass ratio of the $W'$ becomes greater than one ($\Gamma(W')/M_{W'}>1$), signifying the breakdown of perturbation theory. We will exclude this region of the $M_D-M_{W'}$ parameter space from our phenomenological study of heavy quark production in Section VII.

\subsection{Neutral Currents}

We can now calculate the coupling of the fermions to the neutral bosons. All the charged fermions couple to the photon with their standard electric charges. For example,

\begin{equation}
g_{L}^{\gamma uu}=g_{R}^{\gamma uu}=g(e/g)\left(\frac{2}{3}\right)(u_{L}^{0})^{2}+\tilde{g}(e/\tilde{g})\left(\frac{2}{3}\right)(u_{L}^{1})^{2}=\left(\frac{2}{3}\right)e.
\end{equation}

We will be calculating the couplings in the {}``$T_{3}-Q$'' basis.
To do this we use the standard relation between the three quantum
numbers: $Q=T_{3}+Y$. Since the fermions derive their $SU(2)$ charge from more than one site, we will calculate, for example, the $T_{3}$ coupling of two light fields to the $Z$ as $\sum_{i}g_i(f_i)^2 v^{i}_{Z}$. The left handed $Z$ coupling to SM fermions is calculated to be:

\begin{eqnarray}
g_{L}^{Zuu} & = & \left(g_{0}v_{Z}^{0}(u_{L}^{0})^{2}+\tilde{g}v_{Z}^{1}(u_{L}^{1})^{2}\right)T_{3}+g'v_{Z}^{2}\left((u_{L}^{0})^{2}+(u_{L}^{1})^{2}\right)(Q-T_{3})\nonumber \\
 & = & -\frac{e}{\textrm{sin}\,\theta_{w}\,\textrm{cos}\,\theta_{w}}\left(T_{3}-Q\,\textrm{sin}^{2}\,\theta_{w}\right).
\label{ZuuL}
\end{eqnarray}
All the other couplings can be similarly computed and we present the results in Table III.

\begin{table}
\begin{center}
\renewcommand\arraystretch{2}
\resizebox{19cm}{!} {
  \begin{tabular}[b]{| c || c | c | }
    \hline
     Coupling & computed as & Strength \\ \hline\hline
          $g_{L}^{Zuu}$ & $\left(g_{0}v_{Z}^{0}(u_{L}^{0})^{2}+\tilde{g}v_{Z}^{1}(u_{L}^{1})^{2}\right)T_{3}+g'v_{Z}^{2}\left((u_{L}^{0})^{2}+(u_{L}^{1})^{2}\right)(Q-T_{3})$ & $-\frac{e}{\textrm{sin}\,\theta_{w}\,\textrm{cos}\,\theta_{w}}\left(T_{3}-Q\,\textrm{sin}^{2}\,\theta_{w}\right)$ \\ \hline
     $g_{L}^{ZuU}$ & $\left(g_{0}v_{Z}^{0}u_{L}^{0}U_{L}^{0}+\tilde{g}v_{Z}^{1}u_{L}^{0}U_{L}^{1}\right)T_{3}+g'v_{Z}^{2}\left(u_{L}^{0}U_{L}^{0}+u_{L}^{1}U_{L}^{1}\right)(Q-T_{3})$ & $-\frac{e\, x}{2\,\sqrt{2}\,\textrm{sin}\,\theta_{w}\,\textrm{cos}\,\theta_{w}}T_{3}$\\ \hline
     $g_{L}^{ZUU}$ & $\left(g_{0}v_{Z}^{0}(U_{L}^{0})^{2}+\tilde{g}v_{Z}^{1}(U_{L}^{1})^{2}\right)T_{3}+g'v_{Z}^{2}\left((U_{L}^{0})^{2}+(U_{L}^{1})^{2}\right)(Q-T_{3})$ & $-\frac{e}{\textrm{sin}\,\theta_{w}\,\textrm{cos}\,\theta_{w}}\left(\frac{1}{2}\left(1+\frac{x^{2}}{8}(4-\textrm{sec}^{2}\,\theta_{w})\right)T_{3}-Q\,\textrm{sin}^{2}\,\theta_{w}\right)$ \\ \hline
     $g_{R}^{Zuu}$ & $\tilde{g}v_{Z}^{1}(u_{R}^{1})^{2}T_{3}+g'v_{Z}^{2}\left((u_{R}^{1})^{2}+(u_{R}^{2})^{2}\right)(Q-T_{3})$ & $e (Q-T_{3}) \textrm{tan}\theta_{w}$ \\ \hline
     $g_{R}^{ZuU}$ & $\tilde{g}v_{Z}^{1}(u_{R}^{1})(U_{R}^{1})T_{3}+g'v_{Z}^{2}\left((u_{R}^{1})(U_{R}^{1})+(u_{R}^{2})(U_{R}^{2})\right)(Q-T_{3})$ & $0$\\ \hline
     $g_{R}^{ZUU}$ & $\tilde{g}v_{Z}^{1}(U_{R}^{1})^{2}T_{3}+g'v_{Z}^{2}\left((U_{R}^{1})^{2}+(U_{R}^{2})^{2}\right)(Q-T_{3})$ & $-\frac{e}{\textrm{sin}\,\theta_{w}\,\textrm{cos}\,\theta_{w}}\left(\frac{1}{2}(1-\frac{x^{2}}{8})\, T_{3}-Q\,\textrm{sin}^{2}\,\theta_{w}\right)$ \\ \hline
     $g_{L}^{Z'uu}$ & $\left(g_{0}v_{Z'}^{0}(u_{L}^{0})^{2}+\tilde{g}v_{Z'}^{1}(u_{L}^{1})^{2}\right)T_{3}+g'v_{Z'}^{2}\left((u_{L}^{0})^{2}+(u_{L}^{1})^{2}\right)(Q-T_{3})$ & $-\frac{1}{2}\, x\,\textrm{sec}\,\theta_{w}\,\textrm{tan}\,\theta_{w}\,(T_{3}-Q)$\\ \hline
     $g_{L}^{Z'uU}$ & $\left(g_{0}v_{Z'}^{0}U_{L}^{0}u_{L}^{0}+\tilde{g}v_{Z'}^{1}U_{L}^{0}u_{L}^{1}\right)T_{3}$ & $\frac{e}{\sqrt{2}\,\textrm{sin}\,\theta_{w}}\left(1-\frac{x^{2}}{8}\textrm{tan}^{2}\,\theta_{w}\right)T_{3}$  \\ \hline
     $g_{L}^{Z'UU}$ & $\left(g_{0}v_{Z'}^{0}(U_{L}^{0})^{2}+\tilde{g}v_{Z'}^{1}(U_{L}^{1})^{2}\right)T_{3}+g'v_{Z'}^{2}\left((U_{L}^{0})^{2}+(U_{L}^{1})^{2}\right)(Q-T_{3})$ & $-\frac{e}{x\,\textrm{sin}\,\theta_{w}}\left[1-\frac{3x^{2}}{8}(2\,\textrm{sec}^{2}\,\theta_{w}-3\,\textrm{tan}^{2}\,\theta_{w})\right]T_{3}+\frac{1}{2}\, e\, x\,\frac{\textrm{sin}^{2}\,\theta_{w}}{\textrm{cos}\,\theta_{w}}\, Q$  \\ \hline
     $g_{R}^{Z'uu}$ & $\tilde{g}v_{Z'}^{1}(u_{R}^{1})^{2}T_{3}+g'v_{Z}^{2}\left((u_{R}^{1})^{2}+(u_{R}^{2})^{2}\right)(Q-T_{3})$ & $\frac{ex}{2} \textrm{sec}\theta_{w} \textrm{tan}\theta_{w}(Q-T_{3})$ \\ \hline
     $g_{R}^{Z'uU}$ & $\tilde{g}v_{Z'}^{1}(u_{R}^{1})(U_{R}^{1})T_{3}+g'v_{Z'}^{2}\left((u_{R}^{1})(U_{R}^{1})+(u_{R}^{2})(U_{R}^{2})\right)(Q-T_{3})$ & $0$ \\
\hline
     $g_{R}^{Z'UU}$ & $\tilde{g}v_{Z'}^{1}(U_{R}^{1})^{2}T_{3}+g'v_{Z'}^{2}\left((U_{R}^{1})^{2}+(U_{R}^{2})^{2}\right)(Q-T_{3})$ & $-\frac{e}{x\,\textrm{sin}\,\theta_{w}}\left[1-\frac{x^{2}}{8}(2\,\textrm{sec}^{2}\,\theta_{w}-5\,\textrm{tan}^{2}\,\theta_{w})\right]T_{3}+\frac{1}{2}\, e\, x\,\frac{\textrm{sin}^{2}\,\theta_{w}}{\textrm{cos}\,\theta_{w}}\, Q$  \\ \hline

  \end{tabular}
  }
  \caption{The couplings of the light and heavy quarks with the neutral gauge bosons. Ideal fermion delocalization renders the $T_3$ portion of $g^{Z'uu}$ zero while there is a small hypercharge coupling. The coupling of the heavy gauge boson to two heavy quarks, as in the charged current coupling, is seen to be proportional to $1/x$.}
\end{center}
\label{tab:neutral couplings}
\end{table}

While ideal fermion delocalization makes $g_L^{W'ud}$ zero, and likewise makes the $T_3$ portion of $g_L^{Z'uu}$ vanish, there is still a small non-zero hypercharge contribution to $g_L^{Z'uu}$. Also, $g_{L}^{ZuU}$ and $g_{L}^{Z'uU}$ are seen to have only a $T_3$ coupling because the term multiplying $Q-T_{3}$ (hypercharge), $\left(u_{L}^{0}U_{L}^{0}+u_{L}^{1}U_{L}^{1}\right)$, vanishes due to the orthogonality of the fermion wave functions. In the limit $\cos\theta_{w}\rightarrow1$, $g_{L}^{ZuU}$ is seen to correspond exactly to the off diagonal coupling of the $W$, $g_{L}^{WuD}$. As in the case of charged currents, the coupling of two heavy quarks to the $Z'$ is enhanced by a factor $1/x$. This makes $\Gamma(Z')/M_{Z'}>>1$ for small values of $x$.

\section{The Top quark}

The top quark in the model has different properties than the light
quarks since most of its mass is generated by the top Higgs. This section
reviews the masses and eigenstates of the top quark and proceeds to
analyze the delocalization pattern of the top and bottom quarks.

\subsection{Masses and wave functions}

The top quark mass matrix may be read from Eqns. (\ref{top quark mass L}) and (\ref{eqn:Light fermion mass}) and is given by:

\begin{equation}
\left(\begin{array}{cc}
M_{D}\epsilon_{tL} & \lambda_{t}v\textrm{sin}\omega\\
M_{D} & M_{D}\epsilon_{tR}\end{array}\right).
\label{top mass matrix}
\end{equation}

Let us define the parameter 
\begin{equation}
a=\frac{\lambda_{t}\, v\,\textrm{sin}\omega}{M_{D}},
\end{equation}
in terms of which the above matrix can be written as:

\begin{eqnarray}
M_{t}=M_{D}\left(\begin{array}{cc}
\epsilon_{tL} & a\\
1 & \epsilon_{tR}
\end{array}\right).
\label{top mass matrix final}
\end{eqnarray}

Note that we have introduced a left handed delocalization parameter
$\epsilon_{tL}$, that is distinct from the one for the light fermions.
We will see in the next subsection that $\epsilon_{tL}=\epsilon_{L}$
is the preferred value, i.e., the top quark is delocalized in exactly
the same way as the light quarks.

Diagonalizing the top quark mass matrix perturbatively in $\epsilon_{tL}$
and $\epsilon_{tR}$, we can find the light and heavy eigenvalues. The mass of the top quark is:

\begin{equation}
m_{t}= \lambda_{t}v\,\textrm{sin}\,\omega\left[1+\frac{\epsilon_{tL}^{2}+\epsilon_{tR}^{2}+\frac{2}{a}\epsilon_{tL}\epsilon_{tR}}{2(-1+a^{2})}\right].
\label{top mass}
\end{equation}
Thus, we see that $m_{t}$ depends mainly on $v$ and only slightly on $\epsilon_{tR}$, in contrast to the light fermion mass, Eqn. (\ref{eqn:light quark mass}), where the dominant term is $\epsilon_{fR}$ dependent. The mass of the heavy partner of the top is given by:

\begin{equation}
m_{T}=  M_{D}\left[1-\frac{\epsilon_{tL}^{2}+\epsilon_{tR}^{2}+2a\epsilon_{tL}\epsilon_{tR}}{2(-1+a^{2})}\right].
\label{mass of TOP}
\end{equation}
The wave functions of the left and right handed top quark are:

\begin{eqnarray}
t_{L} & = & t_{L}^{0}\psi_{L0}^{t}+t_{L}^{1}\psi_{L1}^{t}\nonumber \\
 & = & \left(1-\frac{\epsilon_{tL}^{2}+a^{2}\epsilon_{tR}^{2}+2a\epsilon_{tL}\epsilon_{tR}}{2(-1+a^{2})^{2}}\right)\psi_{L0}^{t}+\left(\frac{\epsilon_{tL}+a\epsilon_{tR}}{-1+a^{2}}\right)\psi_{L1}^{t}
\label{tL vector}
\end{eqnarray}

\begin{eqnarray}
t_{R} & = & t_{R}^{1}\psi_{R1}^{t}+t_{R}^{2}t_{R2}\nonumber \\
 & = & \left(1-\frac{a^{2}\epsilon_{tL}^{2}+\epsilon_{tR}^{2}+2a\epsilon_{tL}\epsilon_{tR}}{2(-1+a^{2})^{2}}\right)\psi_{R1}^{t}+\left(\frac{a\epsilon_{tL}+\epsilon_{tR}}{-1+a^{2}}\right)t_{R2}.
\label{tR vector}
\end{eqnarray}
The left and right handed heavy top wave functions are the orthogonal
combinations:

\begin{eqnarray}
T_{L} & = & T_{L}^{0}\psi_{L0}^{t}+T_{L}^{1}\psi_{L1}^{t}\nonumber \\
 & = & \left(\frac{\epsilon_{tL}+a\epsilon_{tR}}{-1+a^{2}}\right)\psi_{L0}^{t}+\left(-1+\frac{\epsilon_{tL}^{2}+a^{2}\epsilon_{tR}^{2}+2a\epsilon_{tL}\epsilon_{tR}}{2(-1+a^{2})^{2}}\right)\psi_{L1}^{t}
\label{TL vector}
\end{eqnarray}

\begin{eqnarray}
T_{R} & = & T_{R}^{1}\psi_{R1}^{T}+T_{R}^{2}t_{R2}\nonumber \\
 & = & \left(\frac{a\epsilon_{tL}+\epsilon_{tR}}{-1+a^{2}}\right)\psi_{R1}^{t}+\left(-1+\frac{a^{2}\epsilon_{tL}^{2}+\epsilon_{tR}^{2}+2a\epsilon_{tL}\epsilon_{tR}}{2(-1+a^{2})^{2}}\right)t_{R2}.
\label{TR vector}
\end{eqnarray}

\subsection{$Zb\bar{b}$ and choice of $\epsilon_{tL}$}

Since the $b_L$ is the $SU(2)$ partner of the $t_L$, its delocalization is (to the extent that
$\epsilon_{bR}\simeq 0$) also determined by $\epsilon_{tL}$. Thus, we can compute the tree
level value of the $Zb\bar{_{L}b_{L}}$coupling and use it to constrain
$\epsilon_{tL}$. This coupling is given by:

\begin{eqnarray}
g_{L}^{Zbb} & = & \left(g_{0}v_{Z}^{0}(b{}_{L}^{0})^{2}+g_{1}v_{Z}^{1}(b_{L}^{1})^{2}\right)T_{3}+\tilde{g}v_{Z}^{2}\left((b_{L}^{0})^{2}+(b_{L}^{1})^{2}\right)(Q-T_{3})\nonumber \\
 & = & -\frac{e}{\textrm{sin}\,\theta_{w}\,\textrm{cos}\,\theta_{w}}\left((1+\frac{x^{2}}{4}-\frac{\epsilon_{tL}^{2}}{2})T_{3}-Q\,\textrm{sin}^{2}\,\theta_{w}\right).
\label{ZbbL}
\end{eqnarray}
This exactly corresponds to the tree-level SM value provided that
$\epsilon_{tL}$ satisfies
\begin{equation}
\epsilon_{tL}=\frac{x}{\sqrt{2}}.
\label{top delocalization}
\end{equation}
We see that this matches the delocalization condition for the light quarks, Eqn. (\ref{eqn:IDF}). Thus, we see that the left-handed top quark is to be delocalized in exactly the same way as the light fermions if we are to avoid significant tree-level corrections to the SM $Zb_{L}\bar{b}_{L}$ value. Henceforth, we shall be choosing this value for $\epsilon_{tL}$.

\subsection{$\Delta\rho$ and $M_{D}$}

The contribution of the heavy top-bottom doublet to $\Delta\rho$
can be evaluated in this model and is given by the same expression as in
\cite{three site ref}. It is:
\begin{equation}
\Delta\rho=\frac{M_{D}^{2}\,\epsilon_{tR}^{4}}{16\,\pi^{2}\, v^{2}}.
\label{Delta rho}
\end{equation}
The important difference now is that, since the top quark mass is dominated by the vev of the top Higgs instead of $M_{D}$ (see Eqn. (\ref{top mass})), $\epsilon_{tR}$ can be as small as the $\epsilon_R$ of any light fermion. Thus, there is no conflict between the twin goals of getting a large top quark mass and having an experimentally admissible value of $\Delta\rho$. This enables us to have heavy fermions in this model that are light enough to be seen at the LHC. We explore this in detail in the next section.

\section{Heavy fermion phenomenology at hadron colliders }

We are now prepared to investigate the collider phenomenology of this model. As we have just seen, there is no tension between getting the correct values of the top quark mass and the $\rho$ parameter in this model. Thus, the mass of the heavy quarks do not necessarily lie in the TeV range as in \cite{three site ref}.  The current CDF lower bounds on heavy up-type quarks (decaying via charged currents) and down-type quarks (decaying via neutral currents) are 284 GeV and 270 GeV, respectively, at 95\% C.L. \cite{CDF heavy quark bound}. Thus, in our phenomenological analysis, we will be concentrating
on new quarks whose masses are between 300 GeV and 1 TeV, corresponding to $M_D$ in a similar range.  

An important point to note is that the diagonal coupling of the heavy $W'$ or $Z'$ to two heavy fermions is enhanced by a factor $1/x$ , where $x$ is our small expansion parameter. Thus, if the masses are such that the heavy gauge bosons can decay to two heavy fermions, then we are in a situation where $\Gamma_{W'}/M_{W'}, \Gamma_{Z'}/M_{Z'} >1$, rendering perturbative
analysis invalid. In our analysis of the phenomenology, we will always choose $M_{W',Z'} < {2}M_D$.    We will study both pair and single production channels.

\subsection{Heavy fermion decay}

\begin{figure}[tb]
\begin{center}
\includegraphics[width=1.75in]{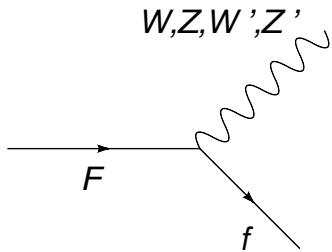}
\caption{The decay modes of the heavy quarks in the theory.
The decay rate is controlled by the off-diagonal left handed coupling
of the vector boson to a heavy fermion and the corresponding light
fermion (the corresponding right handed coupling vanishes in the limit of massless light fermions).
}
\label{fig:Fermion decay}
\end{center}
\end{figure}

The heavy fermions in the model decay to a vector boson and a light fermion. If the heavy fermion is massive enough, the vector boson could be the $W'$ or $Z'$ in the theory as well as the $W$ or $Z$ (Figure \ref{fig:Fermion decay}).\footnote{The situation changes slightly for the heavy top quark, for which decay into top pions is allowed. The study of the top sector of this model is deferred to a future publication.}

In the limit where the mass of the light fermion is zero, the rate of decay to charged gauge bosons (denoted by $V$) is given by:
\begin{eqnarray}
\Gamma=\frac{g_{VFf}^{2}}{32\pi}\frac{M_{D}^{3}}{m_{V}^{2}}\left(1-\frac{m_{V}^{2}}{M_{D}^{2}}\right)^{2}\left(1+2\frac{m_{V}^{2}}{M_{D}^{2}}\right).
\label{decay rate of F}
\end{eqnarray}
\noindent In the limit that the Dirac mass is much higher than the
$W$ and $W'$ boson masses, the terms in the parantheses can be approximated
by $1.$ Thus, we see that the decays into $W$ and $W'$ become equally
important because $g_{WQq}^{2}/m_{W}^{2}\approx g_{W'Qq}^{2}/m_{W'}^{2}$.
This is further illustrated in Figure 3, where we can also see that decays to $Z'$ are generally just slightly less likely than those to $W'$, for any value of $M_D$.

\begin{figure}[tb]
\begin{center}
\includegraphics[width=3.5in]{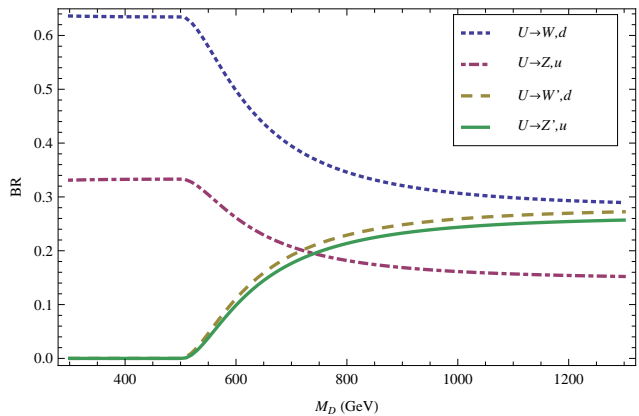}
\caption{The plot of the branching ratio of the heavy quark into the charged and neutral gauge bosons. The masses of the $W'$ and $Z'$ gauge bosons were taken to be 500 GeV.
}
\label{fig:BR}
\end{center}
\end{figure}

\subsection{Heavy quarks at the LHC}

Our goal in this section is to analyze the possible discovery modes of the heavy quarks at the LHC. We will show that it is possible to discover them at 5$\sigma$ level for a large range in the $M_{W'}-M_{D}$ parameter space. We will consider both the (QCD dominated) pair production and the (electroweak) single production of the heavy quarks. Each produced quark immediately decays to either a SM gauge boson plus a light quark or a heavy gauge boson plus a light quark (for $M_{D}>M_{W'}, M_{Z'}$). We will consider the first possiblity in the pair production scenario (Section \ref{Subsection:pair}) and the second in the single production analysis (Section \ref{Subsection:single}) and show that these cover much of the $M_{D}-M_{W'}$ parameter space. For our phenomenological analysis, we used the CalcHEP package \cite{Pukhov-Calchep}.

\subsubsection{Pair production: $pp\rightarrow Q\bar{Q}\rightarrow WZqq\rightarrow lll\nu jj$} \label{Subsection:pair}

We first consider the process $pp\rightarrow Q\bar{Q}$ at the LHC. Pair production of heavy quarks occurs via gluon fusion and quark annihilation processes, shown in Figure 4a. In Figure 4b, we present the production cross section as a function of Dirac mass for a single flavor. We see that the cross-section for the gluon fusion process is higher than that for quark annihilation at low values of $M_{D}$. However, as $M_{D}$ increases, the $q\bar{q}$ channel begins to dominate. This is because the parton distribution function of the gluon falls rapidly with increasing parton momentum fraction.

\begin{figure}[tb]
\centering
\subfigure{
\includegraphics[scale=1.14]{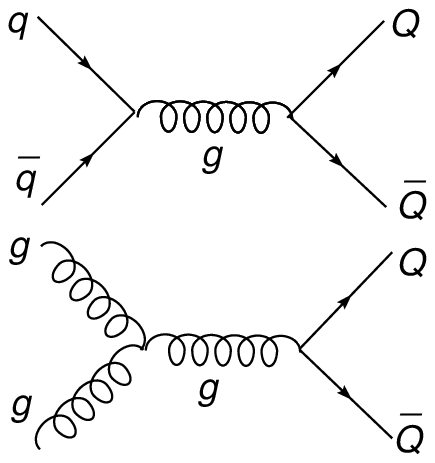}
}
\hspace{14 mm}
\subfigure{
\includegraphics[scale=0.62]{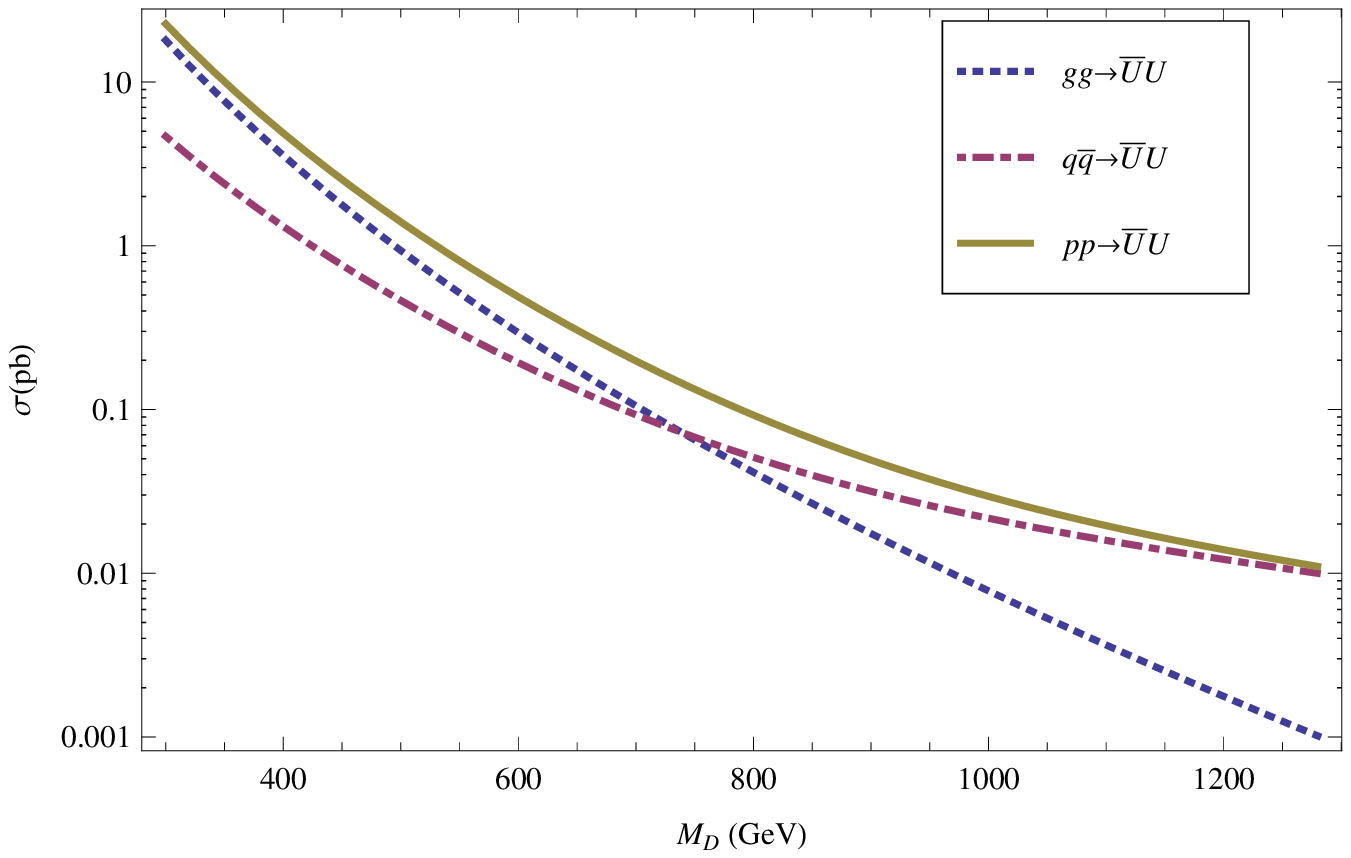}
}
\caption{(a). Pair production of the heavy quarks occurs through
$\bar{q}q$ annihilation and gluon fusion. (b). The cross section for pair production (for one flavor)
as a function of the Dirac mass. As can be seen from the figure, for low values of $M_{D}$, the cross section for the gluon fusion channel is higher than the quark annihilation process. As $M_{D}$ increases, the quark annihilation process begins to dominate because the parton distribution function of the gluon falls rapidly with increasing parton momentum fraction.}
\end{figure}
Each heavy quark decays to a vector boson and a light fermion. For $M_{D}<M_{W',Z'}$, the decay is purely to the standard model gauge bosons. The decay to heavy gauge bosons opens up for $M_{D}>M_{W',Z'}$, and we will analyze this channel while discussing single production of heavy fermions in the next subsection. Here, we look at the signal in the case where one of the heavy quarks decays to a $Z$ and the other decays to a $W$, with the gauge bosons subsequently
decaying leptonically. Thus, the final state is $lll \nu jj$.

\begin{figure}[tb]
\begin{center}
\includegraphics[width=3in]{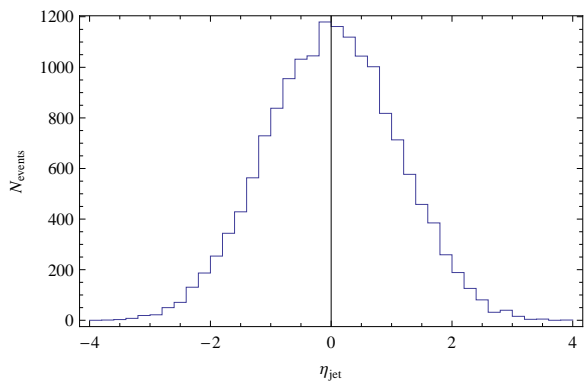}
\caption{The $\eta$ distribution of the outgoing hard jets for the process $pp\rightarrow Q\bar{Q}\rightarrow WZqq\rightarrow lll\nu jj$, corresponding to $M_{D}=700$ GeV and $M_{W'}=500$ GeV for a luminosity
of 100 $fb^{-1}$. One can see that the events are in the central region: $-2.5 < \eta < 2.5$. 
}
\label{fig:rapidity}
\end{center}
\end{figure}

To enhance the signal to background ratio, we have imposed a variety of cuts. We note that the the two jets in the signal should have a high $p_{T}\,$($\sim M_{D}/2)$, since they each come from the 2-body decay of a heavy fermion. Thus, imposing strong $p_{T}$ cuts on the outgoing jets can eliminate much of the SM background without affecting the signal too much. We also expect the $\eta$ distribution of the jets to be largely central (see Figure \ref{fig:rapidity}), which suggests an $\eta$ cut: $|\eta|\leq2.5$. We impose standard separation cuts between the two jets and between jets and leptons to ensure that they are observed as distinct final state particles.   We also  impose basic identification cuts on the leptons and missing transverse energy; the full set of cuts is listed in Table IV.

We study $Q\bar{Q}$ events in which one heavy fermion decays to $W+j$ and the other decays to $Z+j$,  and we further assume that the $W$ and $Z$ decay leptonically.   Since the leptonic decay of $W$ involves neutrinos, it is more convenient to use the $Z+j$ combination as the basis for reconstructing the heavy fermion mass (to avoid the two fold ambiguity in determing momenta when one uses neutrinos).
 We identify the leptons that came from the $Z$ by imposing the invariant mass cut $(M_{Z}-2\, \textrm{GeV})<M_{ll}<(M_{Z}+2\,\textrm{GeV})$.  We then combine this lepton pair with a leading-$p_T$ light jet to reconstruct the heavy fermion mass.   Because one cannot, a priori, tell which light jet came from the $Q$ and which from the $\bar{Q}$, we actually combine the lepton pair first with the light jet of largest $p_T$ and then, separately, with the light jet of next-largest $p_T$, and include both reconstructed versions of each event in our analysis.  

\begin{table}
\begin{center}
  \begin{tabular}{| c | c | }
    \hline
    Kinematic variable & Cut \\ \hline\hline
    $p_{Tj}$ & $>$100 \textrm{GeV} \\ \hline
    $p_{Tl}$ & $>$15 \textrm{GeV} \\ \hline
    \textrm{Missing} $E_{T}$ & $>$15 \textrm{GeV} \\ \hline
    $|\eta_{j}|$ & $<$ 2.5 \\ \hline
    $|\eta_{l}|$ & $<$ 2.5 \\ \hline
    $\Delta R_{jj}$ & $>$0.4 \\ \hline
    $\Delta R_{jl}$ & $>$0.4 \\ \hline
    $M_{ll}$ & 89 \textrm{GeV}$<M_{ll}<93$ \textrm{GeV} \\
    \hline
  \end{tabular}
  \caption{The complete set of cuts employed to enhance the signal to background ratio in the process $pp\rightarrow Q\bar{Q}\rightarrow WZqq\rightarrow lll\nu jj$.  $\Delta R_{jj}=\sqrt{\Delta\eta_{jj} + \Delta \phi_{jj}}$ refers to the separation (in $\eta$-$\phi$ space) between the two jets and, similarly, $\Delta R_{jl}$ refers to the angular separation between a lepton and a jet.}
\end{center}
\label{tab: cuts pair}
\end{table}

When generating the signal events, we included the four flavors\footnote{We do not consider the heavy top and bottom in this analysis. Including them would further enhance the signal, but since the top quark couples to the uneaten top pions, the branching ratios to gauge bosons would be different from that of the heavy partners of the first two generations. We will present the phenomenology of the third generation in a future work).}  of heavy quarks, $U, D, C, S$, that should have similar phenomenology. In Figure 6, we present the invariant mass distribution ($M_{jet + dilepton}$) for events generated assuming $M_{W'}$= 500 GeV and with all the cuts in Table IV imposed; the left-hand (right-hand) panel shows events with $M_D = 300$ GeV ($700$ GeV).   As mentioned earlier, each event appears twice in the plots because one cannot tell which jet came from the $Q$ and which from the $\bar{Q}$ decay.  
This enhances the number of signal events, but also creates the small number of off-peak events in the distributions (Figure 6). We verified that for the $M_{D}$ values of interest, these off-peak events are never numerous enough to compete with the signal; in fact, this can be directly seen from Figure 6.

\begin{figure}[h!]
\centering
\subfigure{
\includegraphics[scale=0.6]{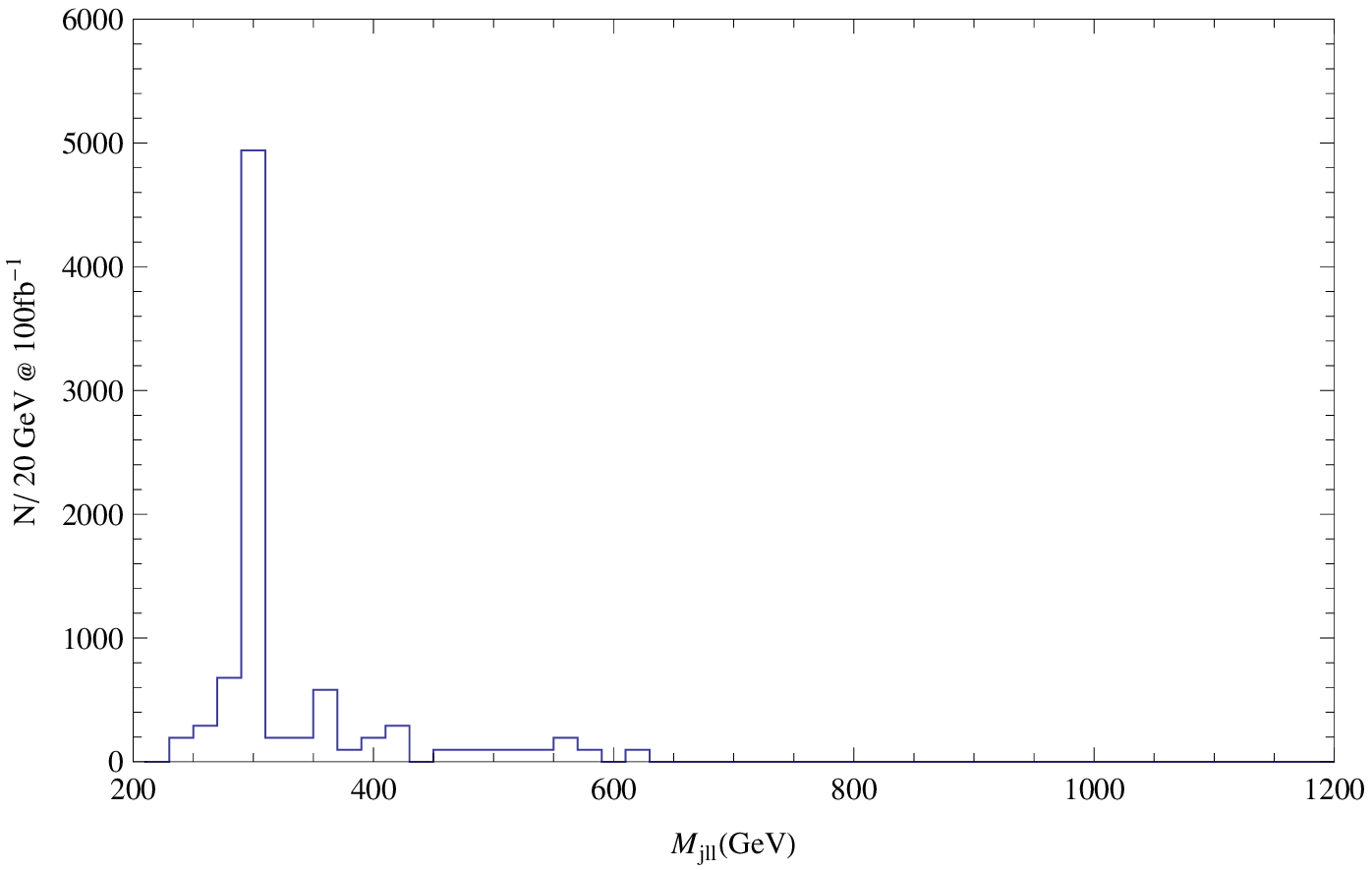}
}
\subfigure{
\includegraphics[scale=0.6]{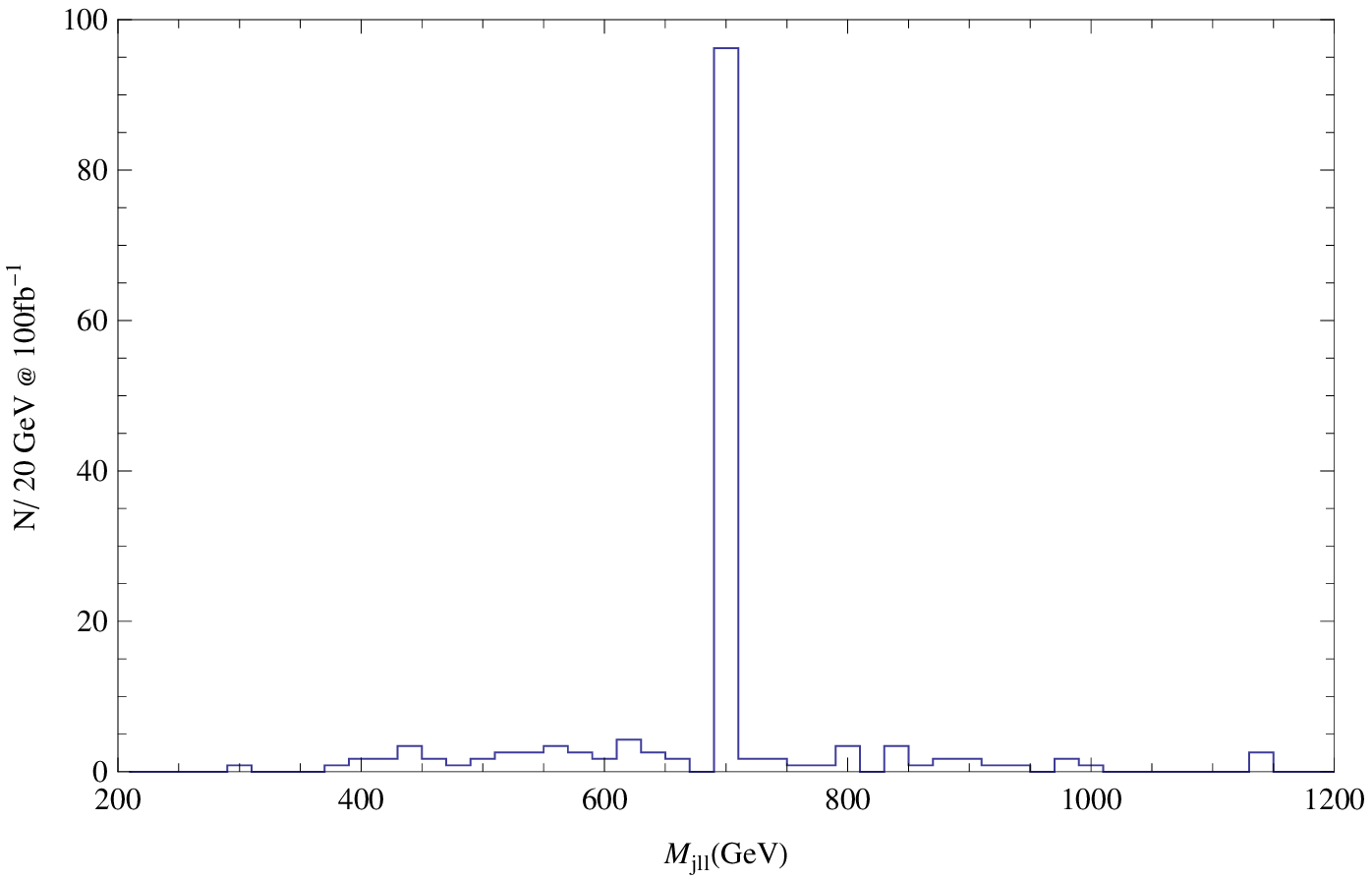}

}
\label{fig:pair distributions}
\caption{Predicted signal invariant mass distributions $M_{llj}$ for $M_D$ = 300 GeV and $M_D$ = 700 GeV for a fixed $M_{W'}$ = 500 GeV. The small number off peak events arises because we added the distributions corresponding to the jets from both $Q$ and $\bar{Q}$ decays, as described in the text.}
\end{figure}

In each of the plots in Figure 6, the signal distribution is clearly seen to peak at the value of $M_{D}$. We estimate the size of the peak by counting the signal events in the invariant mass window:
\begin{equation}
(M_{D}-10\, \textrm{GeV}) <M_{jll}<  (M_{D}+10\, \textrm{GeV}).
\label{Invariant mass cut}
\end{equation}
To analyze the SM background, we fully calculated the irreducible $pp\rightarrow ZWjj$ process and subsequently decayed the $W$ and $Z$ leptonically. Once we imposed all the cuts discussed above on the final state $lll \nu jj$, we find that the cuts entirely
eliminate the background for the range of $M_{D}$ values of interest
to us. The most effective cut for reducing the SM background is the strong $p_{T}$ cut imposed on both the jets.

\begin{figure}[tb]
\begin{center}
\includegraphics[width=3.5in]{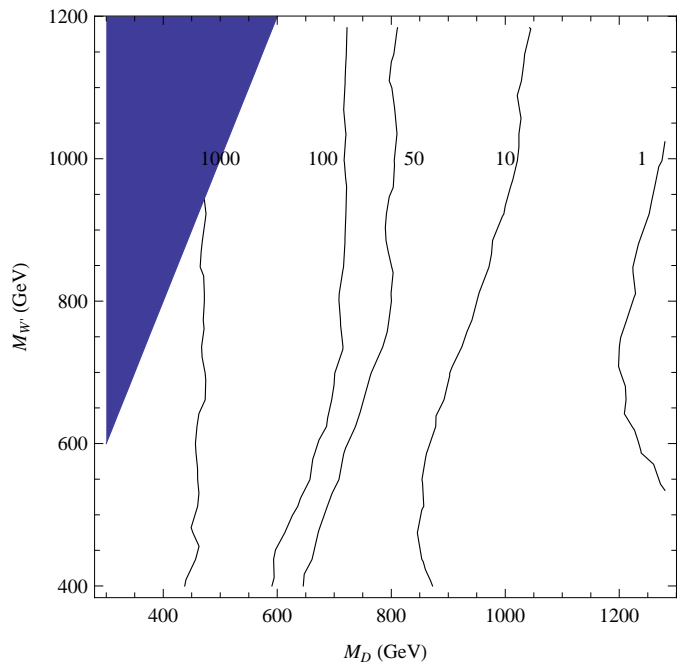}
\caption{Contour plot of number of events in the pair production
case $pp\rightarrow Q\bar{Q}\rightarrow WZqq\rightarrow lll\nu jj$ for a fixed integrated luminosity of 100 $fb^{-1}$. The shaded
region corresponds to the region of non-perturbative $W'$ decays, $M_{W'}>2M_{D}$.
}
\label{fig:pair-signal}
\end{center}
\end{figure}

We find there is an appreciable number of signal events in the region of parameter space where $Q\rightarrow Vq$ decays are allowed but $Q\rightarrow V'q$ decays are kinematically forbidden. The precise number is controlled by the branching ratio of the heavy fermion into the standard model vector bosons. In Figure \ref{fig:pair-signal}, we present a contour plot of the number of expected events in the $M_{D}-M_{W'}$ plane for a fixed luminosity of 100 $fb^{-1}.$

Since the SM background is negligible, if we assume the signal events are Poisson-distributed, then we can take 10 events to represent a 5$\sigma$ signal at 95\% c.l.  (i.e., the minimum number of events required to report discovery). Given that we expect at least 10 signal events over most of the area of the plot, we see that the pair-production process we have studied spans almost the entire parameter space. However, as may be seen from Figure \ref{fig:pair-signal}, in the region where $M_{D}\geq \textrm{900 GeV}$ and $M_{W'}\leq M_{D}$ there will not be enough signal events for the discovery of the heavy quark since the decay channel $Q \to W' q$ becomes significant. In order to explore this region, we will now investigate the single production channel where the heavy quark decays to a heavy gauge boson.

\subsubsection{Single production: $pp\rightarrow Qq\rightarrow W'qq'\rightarrow WZqq'$} \label{Subsection:single}

The single production channel of heavy fermions is electroweak in nature, in contrast to the pair production process considered above.  But the smaller cross sections can be compensated if we exploit the fact that the $u$ and $d$ are valence quarks, and hence their parton distribution functions do not fall as sharply as the gluon's for large parton momentum fraction. Also, there is less phase space suppression in the single production channel than in the pair production case. Thus, we analyze the processes $\left[u,u\rightarrow u,U \right ]$, $\left [d,d\rightarrow d,D  \right ]$ and $\left [u,d\rightarrow u,D \ {\rm or}\ U,d\right ]$. These occur through a $t$ channel exchange of a $Z$ and $Z'$ (Figure 8a). In Figure 8b, we show the cross section of the single production of one flavor of the heavy quark as a function of the Dirac mass. Since we want to look at the region of parameter space where $M_{W'}$ is smaller than $M_{D},$ we let the heavy quark decay to a $W'$. The $W'$ decays 100\% of the time to a $W$ and $Z$, because its coupling to two SM fermions is zero in the limit of ideal fermion delocalization (see Eqn.(\ref{eqn:IDF})). We constrain both the $Z$ and $W$ to decay leptonically so the final state is $lll \nu jj$. 

In principle, one could also consider the case in which the heavy quark decay involves a  $Z'$ rather than a $W'$. The only (small) difference would be that the $Z'$ does not decay to a pair of $W$'s 100\% of the time.   The ideal fermion delocalization condition only makes the $T_3$ coupling of the $Z$ to SM fermions zero, while there is a small non zero hypercharge coupling proportional to $x$.  For the present, we restrict ourselves to $W'$ decays of the heavy quark.

\begin{figure}[tb]
\centering
\subfigure{
\includegraphics[scale=1.14]{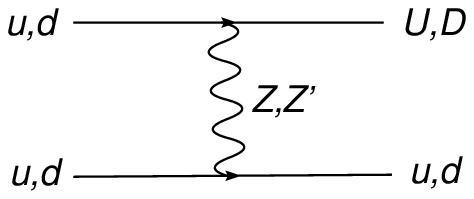}
}
\hspace{14 mm}
\subfigure{
\includegraphics[scale=0.62]{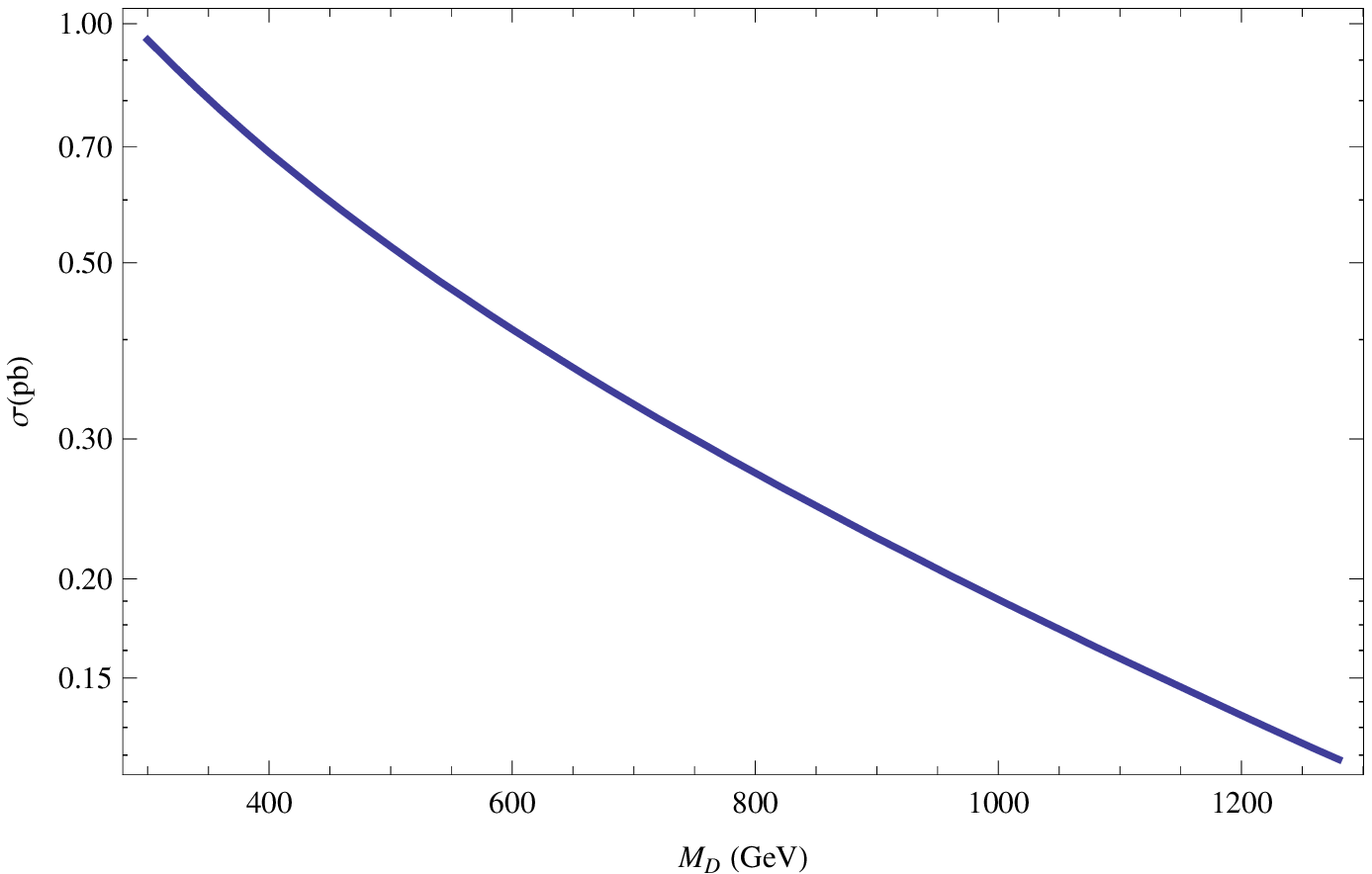}
}
\caption{(a). Feynman diagram for the $t$ channel single production
of the heavy fermion via the exchange of the $Z$ and the $Z'$ bosons. (b). Cross section for the $t$ channel single production
of the heavy fermion as a function of the Dirac mass $M_D$. It is seen to fall more gradually as compared to that of the pair production case. }
\label{fig:single-production}
\end{figure}

As in the case of pair production, we expect the jet from the decay of the heavy quark to have a large $p_{T}$, and hence we will impose a strong $p_{T}$ cut on this ``hard jet". As before, this jet is going to be largely in the central direction and hence one can impose the same $\eta$ cut on the hard jet.  On the other hand, we expect the $\eta$ distribution of the ``soft jet"
arising from the light quark in the production process to be in the forward region, $2<|\eta|<4$. We impose the same $\Delta R$ jet separation and jet-lepton separation cuts as before. We impose basic identification cuts on the leptons and missing transverse energy. The complete set of cuts is shown in Table V.

\begin{table}
\begin{center}
  \begin{tabular}{| c | c | }
    \hline
    Kinematic variable & Cut \\ \hline\hline
   $p_{Tj\ \mbox{hard}}$ & \quad $>$200 \textrm{GeV} \quad \\ \hline
   $p_{Tj\ \mbox{soft}}$ & $>$15 \textrm{GeV} \\ \hline
   $p_{Tl}$ & $>$15 \textrm{GeV} \\ \hline
   Missing $E_{T}$ & $>$15 \textrm{GeV} \\ \hline
   $|\eta_{j\ \mbox{hard}}|$ & $<$ 2.5 \\ \hline
   $|\eta_{j\ \mbox{soft}}|$ & 2$<|\eta|<4$ \\ \hline
   $|\eta_{l}|$ & $<$ 2.5 \\ \hline
   $\Delta R_{jj}$ & $>$0.4 \\ \hline
   $\Delta R_{jl}$ & $>$0.4 \\ \hline
  \end{tabular}
  \caption{\label{tab:cuts-single}The complete set of cuts employed to enhance the signal to background ratio in the process $pp\rightarrow Qq \rightarrow W'q'q \rightarrow WZq'q\rightarrow lll\nu jj$. ``hard" refers to the jet with greater transverse momentum ($p_T$) while ``soft" refers to the jet with smaller transverse momentum ($p_T$).  $\Delta R_{jj}$ refers to the separation (in $\eta$-$\phi$ space) between the two jets and, similarly, $\Delta R_{jl}$ refers to the angular separation between a lepton and a jet.}
\end{center}
\end{table}
 
\begin{figure}[tb]
\centering
\subfigure{
\includegraphics[scale=0.6]{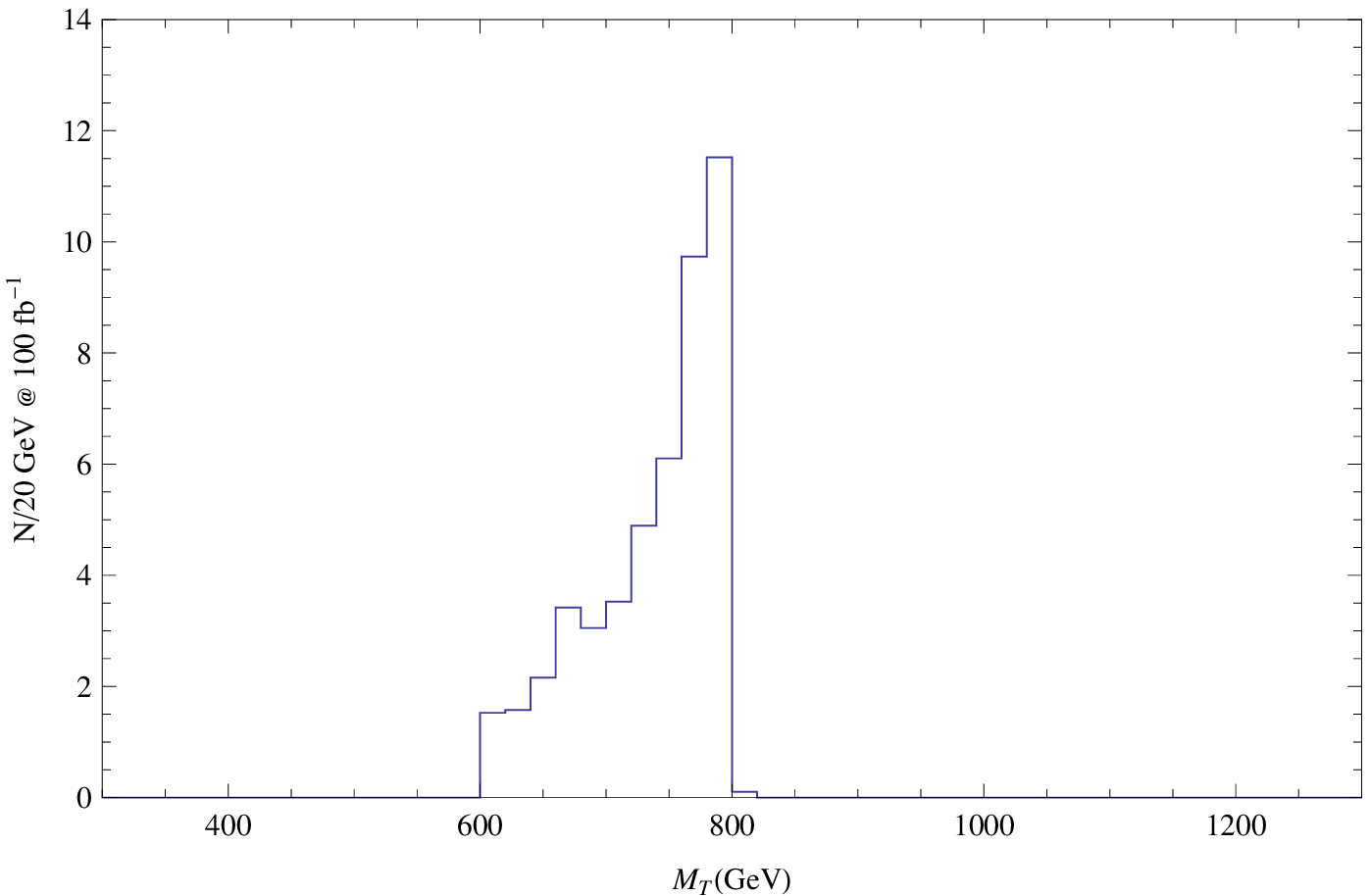}
\label{fig:800500single}
}
\subfigure{
\includegraphics[scale=0.6]{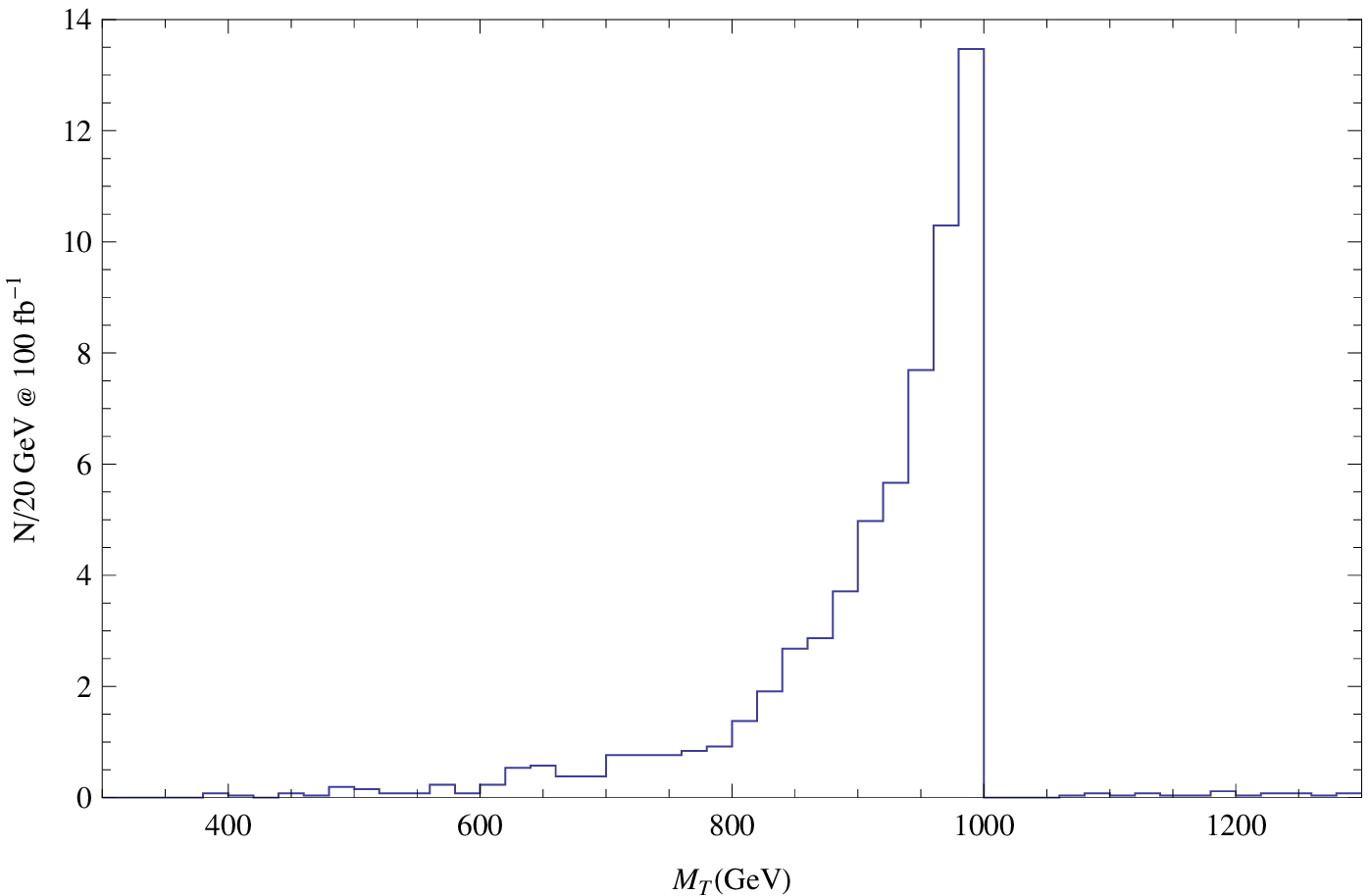}
\label{fig:1000500single}
}
\caption{\label{fig:single distributions}The transverse mass distribution of the signal events for the single production of a heavy quark in the model, for $M_D$ = 800 GeV (left) and 1 TeV (right) and for a fixed $M_{W'}$ = 500 GeV. The cuts are given in Table \ref{tab:cuts-single}.  The bin size is 50 GeV. It is seen that the signal falls sharply at $M_D$.}
\end{figure}

The leptonic $W$ decay introduces the usual two fold ambiguity in determining the neutrino momentum and hence, we have performed a transverse mass analysis of the process, defining the transverse mass variable \cite{Transverse mass definition} of interest as:
\begin{equation}
M_{T}^{2}=\left(\sqrt{M^{2}(lllj)+p_{T}^{2}(lllj)}+\left|p_{T}(missing)\right|\right)^{2}-\left|\overrightarrow{p_{T}}(lllj)+\overrightarrow{p_{T}}(missing)\right|^{2}
\label{Transverse mass variable}
\end{equation}
We expect the distribution to fall sharply at $M_{D}$ in the narrow width approximation, and indeed we find that there are typically few or no events beyond $M_{D}+20$ GeV in the distributions (see Figure \ref{fig:single distributions}). Thus, we take the signal events to be those in the transverse mass window:
\begin{equation}
(M_{D}-200\, \textrm{GeV})  <M_{T}<  (M_{D}+20\, \textrm{GeV}).
\label{Transverse mass cut}
\end{equation}

\begin{figure}[tb]
\begin{center}
\includegraphics[width=3in]{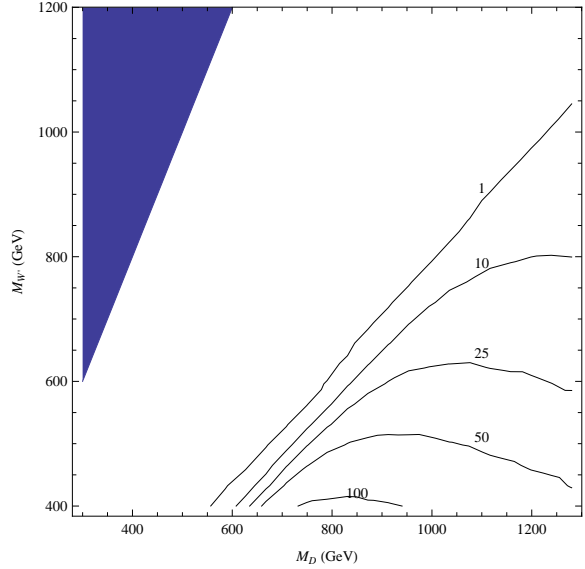}
\caption{Contour plot of the number of signal events for the single
production channel $pp\rightarrow Qq \rightarrow W'q'q \rightarrow WZq'q\rightarrow lll\nu jj$ for an integrated luminosity of 100 $fb^{-1}.$
The shaded region is where $M_{W'}>2M_{D}$ and is non perturbative.
One can see there is a considerable number of events in the low $M_{W'}$
region of the parameter space
}
\label{fig:Single-events}
\end{center}
\end{figure}

We show a contour plot of the number of signal events for an intergrated
luminosity of 100 $fb^{-1}$ in Figure (\ref{fig:Single-events}). It is seen that there are no events
in the $M_{D}<M_{W'}$ region because we require the heavy quarks
to decay to $W'$.
Also, in the region of interest, one can see that there is an appreciable
number of events.

\begin{figure}[tb]
\begin{center}
\includegraphics[width=4in]{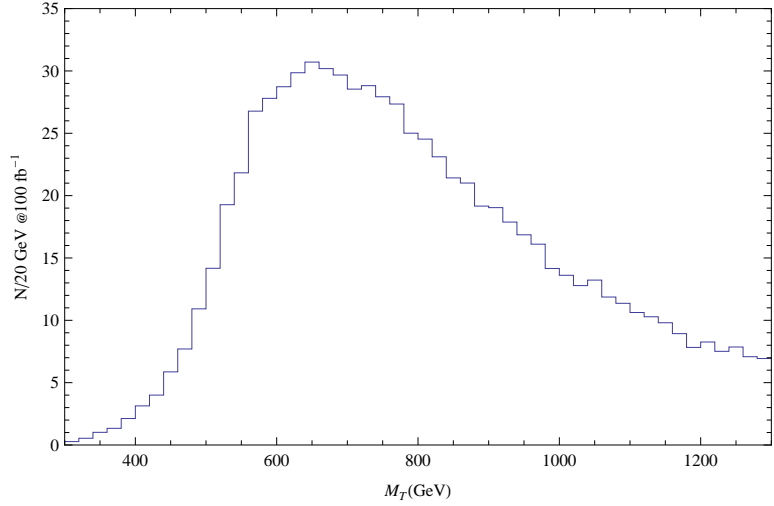}
\caption{The SM background for the single production channel, $pp\rightarrow WZjj \rightarrow jjl\nu ll$, calculated by summing over the $u,$ $d,$ $c$, $s$ and gluon jets and the first two
families of leptons. The bin size is 20 GeV.
}
\label{fig:SM-WZjj}
\end{center}
\end{figure}

The SM background for this process, $pp\rightarrow WZjj \rightarrow jjl\nu ll$, was calculated 
summing over the $u,$ $d,$ $c$, $s$ and gluon jets and the first two
families of leptons. Since we apply a strong $p_{T}$ cut on only one of the jets (unlike in the pair production case), there is a non zero SM background. We show the SM transverse mass distribution in
Figure \ref{fig:SM-WZjj}.

\begin{figure}[tb]
\begin{center}
\includegraphics[width=3in]{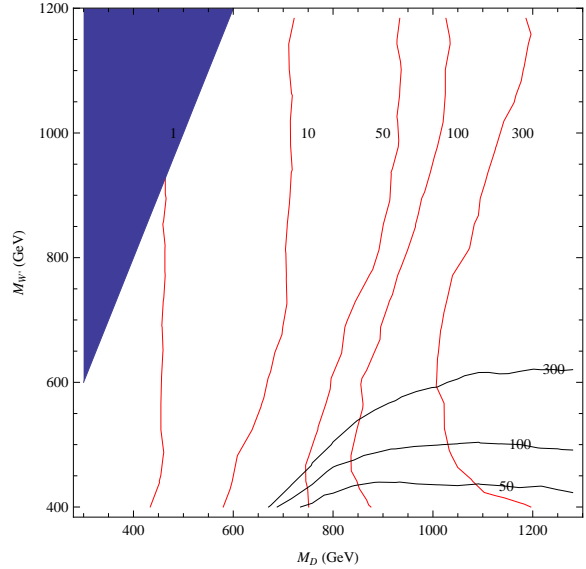}
\caption{Luminosity required for a $5\sigma$ discovery of
the heavy vector fermions at the LHC in the single (blue curves, nearly horizontal) and pair
(red curves, nearly verticle) production channels. The shaded portion is non perturbative
and not included in the study. It is seen that the two channels are
complementary to one another and allow almost the entire region to
be covered in 300 $fb^{-1}.$
}
\label{fig:Luminosity plot}
\end{center}
\end{figure}

The luminosity necessary for a $5\sigma$ discovery at 95\% c.l. can be calculated by requiring $(N_{signal}/\sqrt{N_{bkrnd}}) \geq 5$, as per a Gaussian distribution. It is instructive to look at the results of this analysis by combining it with the previous pair production case, as the two cover the $M_{W'}<M_{D}$ and $M_{W'}>M_{D}$ regions of the $M_{W'}-M_{D}$ parameter space respectively. Thus, we present a combined plot of the required luminosity for a $5\sigma$ discovery of these heavy vector quarks (at 95\% c.l.)  at the LHC in Figure (\ref{fig:Luminosity plot}).

One can see that almost the entire parameter space is covered, with the pair and single production channels nicely complementing each other. Before we conclude, however, we would like to comment briefly on how our analysis compares with other models with vector quarks.

\section{Related Vector Quark Models}

There are many other theories that feature heavy quarks with vector like couplings, as in the present model. In this section, we would like to briefly explain how our phenomenological analysis compares with these. One important feature of deconstructed Higgsless models of the kind discussed in this paper is ideal fermion delocalization, which does not allow the heavy charged gauge bosons in the theory to couple to two standard model fermions. This constrains the $W'$ to decay only to $W$ and $Z$, thus providing a tool to distinguish this class of models from others. There are, however, certain features of this model that are generic, like the vector nature of the heavy quark couplings.

In the context of Little Higgs Models \cite{Little Higgs ref}, there have been studies of the LHC phenomenology of the T-odd heavy quarks \cite{Hubisz paper}. The cross sections for the production of heavy T-quark pairs are comparable to the ones in our study. However, in those models, the heavy T-quark necessarily decays to a heavy photon (due to constraints of conserving T parity). Also, in \cite{Choudhury and Ghosh}, the authors study the pair production of heavy partners of the 1st and 2nd generation quarks in the context of the Littlest Higgs Models \cite{Littlest Higgs 1,Littlest Higgs 2,Littlest Higgs 3,Littlest Higgs 4}. They consider decays exclusively to the heavy gauge bosons in the
theory, which then decay to the standard model gauge bosons plus a heavy photon. Thus, the final state, though still $llljjE\!\!\!\!/\ _{T}$, is kinematically different. In particular, strong cuts on the missing energy are now an important part of the analysis, because part of $E\!\!\!\!/\ _{T}$ is due to the heavy photons. Ref.\cite{Han and Logan} presents a comprehensive study of the production and decay of heavy quarks by separating out the partners of the 3rd generation from the others and analysing them separately. The authors let the heavy quark decay to a SM $W$ boson and a light quark, but in their analysis, they neglect the mass of the $W$ boson compared to its momentum (since it is highly boosted). Thus, when the $W$ decays to a $l\nu$ pair, the direction of the neutrino momentum can be approximated to be parallel to that of the charged lepton, which enables them to recontruct the full neutrino momentum and create an invariant mass peak for the heavy quark (as opposed to a transverse mass analysis).  Clearly, the final states and/or kinematics in all of these analyses differ significantly from those considered in our analysis.

In the context of the three site model, the authors of \cite{three site heavy top production} consider the single production of the heavy top quark. As mentioned before, the heavy top in this model is necessarily around a few TeV's and the paper concludes that the most viable channel for detection at the LHC is the subprocess $qb\rightarrow q'T\rightarrow q'Wb$
with the $W$ decaying leptonically.  We have not yet studied heavy top or top-pion phenomenology in our model.

Ref.~\cite{tao and anu paper} presents a model independent analysis of the discovery prospects of heavy quarks at the Tevatron. The authors write down generic charged and neutral current interactions mixing the heavy and the light fermions and proceed to analyze both the pair and single production of these heavy quarks, with decays to the SM gauge bosons. Understandably, the Tevatron reach is much lower than that of the LHC.

\section{Discussions \& Conclusions}

Higgsless models have emerged as an alternative to the Standard Model in explaining the mechanism of electroweak symmetry breaking. These theories use boundary conditions to break gauge symmetries and can be understood in terms of four dimensional gauge theories via the process of deconstruction. A minimal model along these lines was recently
presented \cite{three site ref}, which employed just three sites. A natural feature of this model was the presence of partners to the SM fermions whose masses were of the order of a few TeV or more, because of the twin constraints of getting the top quark mass right and having the $\rho$ parameter under experimental bounds.

In this paper, we presented a minimal extension of the three site model  that incorporates both Higgsless and top-color mechanisms. The model singles out top quark mass generation as arising from a Yukawa coupling to an effective Top-Higgs, which develops a small vacuum expectation value, while electroweak symmetry breaking results largely from a Higgsless mechanism.  An interesting consequence  is that there is no longer a conflict between the need to obtain a realistically large value for the top quark mass and the need to keep $\Delta\rho$ small enough to conform to experiment.  As a result, in this model it is possible to have additional vector-like quarks in the model that are light enough to be discovered at the LHC, without
affecting the tree level couplings of the three site model too much. 

We encoded the model in CalcHEP and analyzed the phenomenology of the heavy quarks. We first considered pair production ($pp\rightarrow Q\bar{Q}\rightarrow WZjj\rightarrow lll\nu jj$) of these heavy fermions. We found that the 5$\sigma$ reach of the pair production channel was $\approx$ 1 TeV, with the maximum reach in the region where $M_{W'}>M_Q$ where the
heavy quark must decay to a $W$ or $Z$. The single production channel ($pp\rightarrow Qj\rightarrow W'jj\rightarrow WZjj\rightarrow lll\nu jj$) complements this nicely because we can study the decay of the heavy quark to a $W'$, and hence are necessarily in the region $M_{W'}<M_{D}$. By combining both these analyses, we were able to cover most of the $M_{D}-M_{W'}$ parameter space. We conclude that the reach at the LHC for a 5$\sigma$ discovery at 95\% c.l. of the vector quarks in this theory can be as high as 1.2 TeV for an appropriate choice of $M_{W'}$.

Other components of the theory with potentially interesting phenomenology are the heavy partners of the $B$ and $T$ quarks and the uneaten top pions that arise from the extra link in the Moose diagram (Figure 1) and couple preferentially to the third generation. The phenomenology of these states is currently under investigation.

\begin{acknowledgments}
This work was supported in part by the US National Science Foundation under grant PHY-0354226. BC acknowledges Chuan-Ren Chen for useful discussions during the initial stages of this project. NDC would like to thank James Linnemann and James Kraus for helpful discussions during the final stages of this project. RSC and EHS gratefully 
acknowledge the hospitality and support of the  Aspen Center for Physics 
where some of this work was completed.  All the Feynman diagrams
in this paper were generated using JaxoDraw \cite{jaxodraw 1,jaxodraw 2}.

\end{acknowledgments}

\end{document}